\documentclass[aps,12pt,prb,floats,showpacs,eqsecnum,]{revtex4}
\usepackage{graphicx}
\usepackage{amssymb}

\begin{document}

\title{Vibrational instability, two-level systems and Boson peak in
glasses}

\author{D. A. Parshin}
\affiliation{Saint Petersburg State Technical University, 195251 Saint Petersburg, Russia}
\affiliation{Max-Plank-Institut f\"ur Physik komplexer Systeme, D-01187 Dresden, Germany}
\author{H. R. Schober}
\affiliation{Institut f\"ur Festk\"orperforschung, Forschungszentrum
J\"ulich, D-52425, J\"ulich, Germany}
\author{V. L. Gurevich}
\affiliation{A. F. Ioffe Institute, 194021 Saint Petersburg, Russia}
\date{\today}

\begin{abstract}
\baselineskip=2.5ex
  We show that the same physical mechanism is fundamental for two
  seemingly different phenomena such as the formation of two-level systems
  in glasses and the Boson peak in the reduced density of
  low-frequency vibrational states $g(\omega)/\omega^2$. This
  mechanism is the {\em vibrational instability} of  weakly interacting
  harmonic modes. Below some frequency $\omega_c\ll\omega_0$ 
  (where $\omega_0$ is of the order of Debye frequency) 
  the instability, controlled by the anharmonicity,
  creates a new stable {\em universal} spectrum of harmonic vibrations with
  a Boson peak feature as well as double-well potentials with a wide
  distribution of barrier heights. Both are determined by the strength of
  the interaction $I\propto\omega_c$ between the oscillators. 
  Our theory predicts in a natural way a small
  value for the important dimensionless parameter 
  $C=\overline{P} \gamma^2/ \rho
  v^2\approx 10^{-4}$ for two-level systems in glasses.
  We show that $C\approx (W/\hbar\omega_c)^3\propto I^{-3}$ 
  and {\em decreases} with increasing of the interaction strength $I$. The 
  energy $W$ is an important characteristic energy in glasses and is of the 
  order of a few Kelvin. 
  This formula relates the TLS's parameter $C$ with the width of the vibration 
  instability region $\omega_c$ which is typically larger or of the order of the 
  Boson peak frequency $\omega_b$. Since $\hbar\omega_c\gtrsim\hbar\omega_b\gg W$ 
  the typical value of $C$ and therefore 
  the number of active two-level systems is very small, less than one per
  ten million of oscillators, in a good agreement with experiment.
  Within the unified approach developed in the present paper the density
  of the tunneling states and the density of vibrational states at the
  Boson peak frequency are interrelated.
  \end{abstract}
  \maketitle

\baselineskip=2.5ex
\section{introduction}

One of the most typical low-temperature properties of glasses is the
existence of two-level systems (TLS's) which determine at low
temperatures, typically below a few Kelvin, and frequencies less 
than 1~GHz, phenomena such as specific heat,
thermal conductivity, ultrasonic and microwave absorption and many
others (such as the echo, etc)\cite{phillips:81}. Another remarkable
universal property of almost all glasses is the Boson peak in the low
frequency inelastic scattering, proportional to the reduced density
of the vibrational states $g(\omega)/\omega^2$. Compared to the TLS's
it is observed at much higher frequencies, between 0.5 and 2 THz, and
persists to higher temperatures, sometimes up to the glass transition
temperature $T_g$. Usually these two important glassy features are
considered separately and no definite connection between them has
been established so far.

The purpose of the present paper is to propose a physical picture in
which these two seemingly different phenomena are closely
interrelated. We will show that the formation of the Boson peak in
the reduced vibrational density of states (DOS) inevitably leads to the
creation of two-level systems (and vice versa).
A mechanism of the Boson peak formation, implying also the formation
of TLS's was proposed in our recent papers~\cite{GPS,GPS1}. This
mechanism was based on the phenomenon of {\em vibrational
instability} of weakly interacting harmonic oscillators (HO's). The instability 
takes place in the low frequency region, $0 \le\omega \lesssim \omega_c$,
below some characteristic frequency $\omega_c$ which is proportional 
to the interaction strength $I$. Changing the interaction strength 
one can vary the width of the instability region, the position of the Boson peak
and the number of TLS's. The weakness of the interaction $I$ implies that 
$\omega_c\ll\omega_0$, where $\omega_0$ is of the order of Debye 
frequency. Thus the vibrational instability occurs far below the Debye 
frequency $\omega_0$ and has little influence on the high frequency vibrations.

The low-frequency harmonic oscillators we are speaking about 
are realized in glasses as {\em quasi-local} vibrations
(QLV) which can be understood as {\em local} low frequency
vibrations bilinearly coupled to the sound waves~\cite{MMWI,SR:04}.
The existence of quasi-local vibrations in glasses has been confirmed
in a number of papers (see e.g. the literature cited in
Ref.~\cite{GPS}). The local low frequency vibrations are the cores of
the QLV's. The cores represent collective low frequency vibrations of
small groups of atoms~\cite{BGGS,LS:91}. If one plots the potential
energy against the amplitude of one of these modes
\cite{SO:96,LMNS:00}, one obtains a mode potential, as treated in the
soft potential model~\cite{KKI,IKP}. The vibrational instability
results from the weak interaction $I$ of these soft oscillators with
high-frequency oscillators (with $\omega\simeq\omega_0$). 
As a result of this harmonic instability
and the anharmonicity of the glass the initial vibrational density
of states $g_0(\omega)$ is reconstructed in the low frequency 
region $0\le\omega\lesssim \omega_c$ where instability takes place 
and shows the {\em universal} Boson peak feature.

The microscopic origin of these high frequency oscillators, whose
frequencies are much higher than the Boson peak one, is not important
in this picture. It has been shown that, in general, the frequency of
the sound waves can pass through the Ioffe-Regel limit near the Boson
peak frequency~\cite{GPPS,SO:96,ParLaer,S:03,French}. Therefore, the
higher frequency modes might have a rather complex structure. They interact
with the soft oscillators. This interaction causes the vibrational
instability and hence the Boson peak and TLS's formation.

The vibrational instability is a rather general phenomenon and occurs 
in any system of bilinearly coupled harmonic oscillators. 
It can be considered in a purely harmonic approximation.
For example, a system of two oscillators
with masses $M_{1,2}$ and frequencies $\omega_{1,2}$ becomes unstable
if the interaction $I$ between the oscillators exceeds a critical
value~\cite{GPS} $I_c=\omega_1\omega_2\sqrt{M_1M_2}$. If one of two 
frequencies is small then the critical interaction $I_c$ is also small.
In our case we have such an instability due to the
interaction of low and high frequency oscillators.
Therefore, we can simplify the picture to treat it analytically.
The physical reason for the instability in this case is the fact that the high
frequency modes adiabatically follow the motion of the low-frequency
oscillator (a description of this fact can be given within the {\em
adiabatic approximation}). The squared frequency of the soft
oscillator is reduced by an amount proportional to the strength of the
interaction squared and, therefore, even can turn negative. For the example 
of two interacting oscillators with frequencies $\omega_1\ll\omega_2$ 
the new frequency $\widetilde{\omega}_1$ of the low-frequency oscillator 
is given by~\cite{GPS} 
\begin{equation}
\widetilde{\omega}_1^2 = \omega_1^2\left(1-I^2/I_c^2 \right) .
\label{s5cv}
\end{equation}
It is important that it decreases to zero {\em linearly} with $I_c-I$  
($\widetilde{\omega}_1^2\propto I_c-I$, when $I\to I_c$).

This is the essence of the vibrational instability phenomenon in the general case. 
As a result, switching on the interaction between the oscillators, 
we shift {\em all} low frequency modes with $\omega\lesssim\omega_c$
towards the boundary point $\omega=0$ and some of them will cross this 
point (so that the corresponding $\omega^2$ turns negative). 
Therefore, in such a case (if there is no hard gap around zero in the 
initial density of states, $g_0(\omega)$) 
we will have a {\em constant} distribution of renormalized 
$\omega^2$ around $\omega=0$. From that we immediately 
get the right wing of the Boson peak. Indeed, the local anharmonicity 
does not change this important property of uniform distribution of 
renormalized $\omega^2$ around zero. One can show~\cite{GPS} that it restores
the mechanical stability of the system by simply {\em reflecting}
all the negative $\omega^2$ values back to the positive 
$\omega^2$ range  (like in a mirror but with additional stretching 
factor of 2, which is obviously not important, see Eq.~(\ref{1})). 
The strength of the anharmonicity itself plays no role in this 
{\em mirror transformation}.
Now a constant distribution of $\omega^2$ around zero (on the positive side) 
obviously leads to a universal {\em linear law} for the density of states, 
$g(\omega)\propto\omega$ in the interval 
$0\le\omega\lesssim\omega_c$ (independent of the initial DOS 
$g_0(\omega)$). And in its turn 
this linear $\omega$ dependence of the reconstructed DOS just gives 
us the right wing of the Boson peak, 
since $g(\omega)/\omega^2\propto 1/\omega$ and this dependence 
is also universal and independent of the initial DOS 
$g_0(\omega)$. 

We want to stress that this DOS transformation (due to the phenomenon of vibrational 
instability) is rather general and universal since any 
monotonous "traffic" of $\omega^2$ from positive to negative values (due to interaction 
between the oscillators, or due to  changing the temperature or pressure, etc.) 
always gives a constant distribution of $\omega^2$ 
around zero. Therefore, the universal linear DOS, 
$g(\omega)\propto\omega$, and the corresponding universal right wing of the Boson peak, 
$g(\omega)/\omega^2\propto 1/\omega$, inevitably emerge as a result of this instability.

If the origin of the right wing of the Boson peak looks somehow natural, 
the left wing of the Boson peak appears as a result of  the less obvious 
additional transformation of the linear DOS at smaller frequencies.
The point is that when the anharmonicity restores the 
mechanical stability of the system,
single-well potentials describing the unstable 
soft modes with negative $\omega^2$ 
are replaced by double-well ones. This means that the
effective potential energy of the glass in direction of the local
soft mode has two minima separated by a rather low barrier. 
Thus, in this scheme, the two-level systems are created simultaneously with
the Boson peak due to the same mechanism of vibrational instability.
Besides their own high importance (the TLS's physics in glasses and beyond) 
these double-well potentials play also an important role in building the 
left wing of the Boson peak. It can be explained as follows.  

Due to bilinear interaction between the oscillators, $I_{ij}x_ix_j$,  
double-well potentials, with a particle vibrating in one of the wells and therefore 
having a non-zero average displacement $x_{i0}\ne 0$, create random {\em static} 
forces $f_i\approx I_{ij}x_{j0}$ acting on other oscillators. In a purely harmonic case, 
these linear forces would not affect the frequencies (and linear density of states 
$g(\omega)\propto\omega$ would not change). However, together with the local 
anharmonicity the static forces create a universal {\em soft gap}, $g(\omega)\propto\omega^4$,
in the linear density of states. This soft gap is a manifestation of the sea-gull 
singularity~\cite{IKP} 
(see also Ref.~\cite{BGGS}) predicted in the framework of the soft potential 
model for glasses. Recently it was shown that $\omega^4$ 
behavior of the DOS is indeed a universal
feature in disordered systems for low frequency bosonic excitations which are not 
Goldstone modes~\cite{GC}. 

The physical reason for this gap is very transparent. 
Due to anharmonicity there is always a {\em blue shift} of the soft oscillator 
frequency under the action of the static force $f$. For small $\omega$ 
this shift is proportional to $|f|^{1/3}$ and it is anomalously large 
for small forces. The density of states in the gap then can be estimated as follows
\begin{equation}
g(\omega) \propto \int\limits_0^\omega \omega_1 d\omega_1
\int\limits_{-\infty}^{+\infty}\delta\left(\omega - af^{1/3}\right) df
\propto \omega^4 .
\label{12gh}
\end{equation}
As a result, the random static forces (together with anharmonicity) effectively "push out" 
oscillators from the low frequency range to higher frequencies, creating the universal soft gap, $g(\omega)\propto\omega^4$. One can also see this gap  
in the context of mechanical stability of the system but of another kind.
Due to local anharmonicity small frequencies cannot survive in the system 
in the presence of random static forces. In some sense they are not stable 
even though random forces cannot transform single-well potentials into double-well ones.

The width of the $\omega^4$-gap is of the order of the 
Boson peak frequency $\omega_b\propto (\delta f)^{1/3}$, where $\delta f$ is
the width of the random force distribution $P(f)$. 
The Boson peak frequency $\omega_b$ is typically smaller~\cite{foot1} than 
the characteristic frequency $\omega_c$ determining the width of the vibrational 
instability region~\cite{GPS}
\begin{equation}
\omega_b\approx\omega_c\left[\frac{g_0(\omega_c)}{g_0(\omega_0)}
\right]^{1/3} .
\label{xt56}
\end{equation}
The strong inequality $\omega_b\ll\omega_c$ occurs if 
$g_0(\omega_c)\ll g_0(\omega_0)$ i.e. $g_0(\omega)$ is an increasing function 
of $\omega$ (since $\omega_c\ll\omega_0$). 
As a result, $g(\omega)\propto\omega^4$ in the interval 
$0\le\omega\lesssim\omega_b$. Frequencies higher than  
$\omega_b$ are only weakly affected by the static 
forces. Therefore, the linear DOS, 
$g(\omega)\propto\omega$ (and the right wing of the Boson peak) 
which was created in the course of the vibrational instability is conserved
in the interval $\omega_b\lesssim\omega\lesssim\omega_c$. 
Since at small frequencies $\omega\lesssim\omega_b$ we have the universal 
$\omega^4$ gap in the density of states, the left wing of the 
Boson peak takes also the universal form,
$g(\omega)/\omega^2\propto\omega^2$. As a result in the region of the
vibrational instability, $0\le\omega\lesssim\omega_c$, 
we have a universal behavior of the density of states $g(\omega)$ 
($g(\omega)\propto\omega^4$ for $0\le\omega\lesssim\omega_b$ and $g(\omega)\propto\omega$ for
$\omega_b\lesssim\omega\lesssim\omega_c$) with the 
Boson peak feature, independent of the initial DOS $g_0(\omega)$. At higher frequencies, 
above $\omega_c$, we keep the initial DOS $g_0(\omega)$ almost undistorted. 

The Boson peak was the main topic of our previous
papers~\cite{GPS,GPS1}. In the present work we shall concentrate on
the two-level systems, i.e. the level splittings due to the tunneling
through the barriers separating the two minima of the two-well
potentials. But this consideration is not independent of the Boson peak since
we will see that the main parameters of TLS's will be strongly interrelated with
the parameters of the Boson peak. Therefore, these two universal phenomena should be 
considered together.

In the standard tunneling model the TLS's are often characterized by
the so-called dimensionless tunneling strength $C$, Ref.~\cite{Pohl}
\begin{equation}
C=\frac{\overline{P}\gamma^2}{\rho v^2}
\label{yu6g}
\end{equation}
where $\overline{P}$ is the density of states of the TLS's, $\gamma$ the
deformation potential, $\rho$ the mass density of the glass
and $v$ is the average sound velocity. The experimental value of
$C$ for different glasses is small and varies in a narrow  band 
between 10$^{-3}$ and 10$^{-4}$. In our theory such small 
numerical values for $C$ will emerge in a natural way.

Several authors~\cite{Klein,YuLeg,CCYu,Coop} proposed that
the approximate universality and smallness of $C$ in glasses may be a
consequence of the interaction between the TLS's. Roughly speaking the idea
was based on a mean-field approximation. The $i$th TLS produces at a
distance $r_i$ a deformation 
\begin{equation}
\varepsilon_i\simeq \frac{\gamma_i}{\rho v^2r_i^3} ,
\label{w7hy}
\end{equation}
where $\gamma_i$ is the deformation potential of the {\it i}th TLS.
As the deformation is inversely proportional to $r_i^3$
the distribution function of the deformations in a glass 
is a Lorentzian with width $\delta\varepsilon$
proportional to the total concentration $N$ of the TLS's:
\begin{equation}
\delta\varepsilon\simeq\frac{\gamma N}{\rho v^2} .
\label{sgr5}
\end{equation}
The energy $E_i$ (the interlevel spacing) of each TLS changes
under the deformation $\varepsilon$ as
\begin{equation}
\delta E_i = \gamma_i\varepsilon 
\label{q8ngh}
\end{equation}
and from  Eq.~(\ref{sgr5}) one sees that the energies $E_i$ of
the TLS's are distributed in the interval $\delta E$:
\begin{equation}
\delta E \simeq \frac{\gamma^2N}{\rho v^2} .
\label{s5gt}
\end{equation}
For small energies, the density of states $n(E)$ is
independent of both the energy and the concentration of TLS's:
\begin{equation}
n(E) \simeq \frac{N}{\delta E} \simeq \frac{\rho v^2}{\gamma^2} .
\label{s88ik}
\end{equation}

This is the result of a purely classical approach. In this approach the
dimensionless parameter $C_{\rm cl}$
\begin{equation}
C_{\rm cl}\simeq \frac{n(E)\gamma^2}{\rho v^2} \simeq 1
\label{eq}
\end{equation}
is of order unity rather than of the order of
$10^{-4}-10^{-3}$. This was the main difficulty of the theory
outlined in Ref's.~\cite{Klein,YuLeg,CCYu,Coop}. However, if one
accounts for the quantum nature of tunneling, the
situation is improved and the value of $C$ is reduced strongly.
To explain this on the qualitative level 
we remind that the energy $E$ of a TLS consists of two
contributions, i. e. the classical asymmetry $\Delta$ and the quantum
tunneling amplitude $\Delta_0$: $E=\sqrt{\Delta^2+\Delta_0^2}$. So far
we have disregarded the latter.  

According to the standard tunneling model
\begin{equation}
\Delta_0 = \hbar\omega_0\exp(-\lambda),
\label{78uj}
\end{equation}
where
$\lambda$ is the tunneling parameter, distributed uniformly in the
interval $\lambda_{\rm min} < \lambda < \lambda_{\rm max}$. Usually
$\lambda_{\rm min}$ is taken to be about unity. Only TLS's with
$\lambda\simeq 1$ can  tunnel during typical experimental times. 
If $\lambda_{\rm max}\gg 1$, systems with
$\lambda\simeq\lambda_{\rm max}$ cannot tunnel and do not contribute to
the observable properties. Therefore, the relative number of TLS's
participating in the tunneling is proportional to the small number
$1/\lambda_{\rm max}$. As a result, we estimate the
dimensionless parameter $C$ in glasses as
\begin{equation}
C\simeq C_{\rm cl}/\lambda_{\rm max}\simeq 1/\lambda_{\rm max}.
\label{yuh5}
\end{equation}
If for example $\lambda_{\rm max}\simeq 10^{3}$ the dimensionless
parameter $C\simeq 10^{-3}$. Thus the smallness of the dimensionless 
parameter $C$ in our theory is related to typically large values of the 
tunneling parameter $\lambda_{\rm max}$ (and to typically 
high barriers in the system). We will discuss this point 
in Section~\ref{TLS}.

In the same Section we will show that two important parameters,
namely $C$ for TLS's and the characteristic frequency $\omega_c$ for HO's, 
marking the onset of the vibrational instability  
are closely interrelated:
\begin{equation}
C \approx \left(\frac{W}{\hbar\omega_c}\right)^3 .
\label{a45d}
\end{equation}
Here $W$ is an important characteristic energy in glasses~\cite{DPR}. 
Typically it is of the order of a few Kelvin. It determines 
for example the position of the minimum in the reduced 
specific heat~\cite{IKP,BGGS}, $C(T)/T^3$ ($W\approx 2 T_{\rm min}$) 
and some other properties of glasses above one Kelvin~\cite{RB}.
For vitreous silica $W\approx 4$\,K. In particular it follows from this formula 
that the larger is the interaction $I\propto\omega_c$ between the oscillators
the smaller is the TLS's parameter $C$: $C\propto 1/I^3$. And as we will 
see in Section~\ref{TLS} the smaller will be also the density of tunneling states, 
$\overline{P}\propto 1/I^4$. It naturally explains the very old puzzle 
in the physics of glasses, why the number of two-level systems is so small 
(one two-level system for a million of atoms).

At a first glance this interesting result seems to be rather contradicting. 
The stronger the interaction $I$ between the oscillators, the 
larger is the width of the vibrational instability region 
$\omega_c\propto I$ and therefore the higher is the number of double-well
potentials created in the course of stabilization of the system 
due to anharmonicity. The explanation of this seeming paradox is that majority of the 
double-well potentials created due to vibrational instability 
have so high barriers $V$ that they cannot participate
in tunneling at all. As a result only a very small part of the double-well potentials
contributes to the tunneling density of states, $\overline{P}$. 

Since the experimental values of $W$ and $C$ are well known for many glasses 
one can estimate from Eq.~(\ref{a45d}) the important 
characteristic energy $\hbar\omega_c$
which gives the width of the vibrational instability region in glasses
\begin{equation}
\hbar\omega_c\approx WC^{-1/3} .
\label{s6e4}
\end{equation}
For example for a-SiO$_2$ $W=4$\, K and $C=3\cdot 10^{-4}$ giving
$\hbar\omega_c\approx 60$\,K. This falls into the Boson peak region 
($\hbar\omega_b\approx 70$\,K). As a result we see that indeed the Boson peak is placed in
the vibrational instability range. 

\section{Vibrational instability}

To illustrate the idea of a vibrational instability, we consider a
cluster containing a low-frequency harmonic oscillator (HO) with
frequency $\omega_1$ surrounded by a large number, $s-1$, of HO's with
much higher frequencies $\omega_j$ of the order of $\omega_0\gg\omega_1$.
Here $\omega_0$ is a order of magnitude estimate of the high
frequencies. In glasses it usually is of the order of the Debye
frequency. Let $n_0$ be the total concentration of the HO's in the
cluster and $g_0(\omega)$ the normalized initial density of states
(DOS), i. e. the DOS of the HO neglecting their interaction,
\begin{equation}
\label{ao7}
g_0(\omega)=\frac{1}{s}\sum_{i=1}^s\delta(\omega-\omega_i) .
\end{equation}

Including the interaction between the HO's, the total potential
energy of the cluster is
\begin{equation}
U_{\rm tot}(x_1,x_2,...,x_s)=\sum_i\frac{k_i}{2}x_i^2 -
\frac{1}{2}\sum_{i,j\ne i}I_{ij} x_ix_j +
\frac{1}{4}\sum_i A_ix_i^4, \qquad A_i>0 .
\label{potgen}
\end{equation}
Here $x_i$ are the generalized coordinates describing the vibrations
of HO's, $k_i>0$ are the quasielastic constants of noninteracting oscillators 
and $I_{ij}$ determines the bilinear interaction between the
oscillators. To stabilize the system we have added in this equation
the anharmonic terms, $A_ix_i^4$ (with $A_i>0$). The interaction strength
is given by~\cite{GPPS}
\begin{equation}
I_{ij}=g_{ij}J/r^3_{ij},\qquad J\equiv \Lambda^2/\rho v^2
\label{hto1}
\end{equation}
where $g_{ij}\simeq \pm 1$ accounts for the relative orientation of the
HO's, $r_{ij}$ is the distance between HO's, $\rho$ is the mass density
of the glass and $v$ is the  sound velocity.

The interaction between the HO's is due to the coupling between a single
HO and the surrounding elastic medium (the glass). This HO-phonon
coupling has the form~\cite{BGGPRS}
\begin{equation}
\label{int1}
{\cal H}_{\rm int} = \Lambda x\varepsilon ,
\end{equation}
where $\Lambda$ is the coupling constant and $\varepsilon$ is the
strain. Introducing the masses of oscillators $M_i$ we have for the
bare frequencies (neglecting the bilinear interaction) as usual
\begin{equation}
\label{nyt}
\omega_{i}= \sqrt{k_i/M_i} .
\end{equation}
These bare frequencies enter Eq.~(\ref{ao7}) for the initial DOS 
$g_0(\omega)$.
We will assume the characteristic strength of the bilinear
interaction $I\simeq Jn_0$ between the oscillators to be {\em much smaller}
than the typical quasielastic constants, so that
\begin{equation}
  \label{tb2}
 |I|\ll M\omega_0^2\equiv k_0\simeq k_j, \qquad (j\ne 1) ,
\end{equation}
where $M$ is the typical mass of the HO's.

The equation of motion for the low-frequency oscillator is
\begin{equation}
  \label{ss3}
 M_1\ddot x_1 = -k_1x_1+\sum_{j\ne 1}I_{1j}x_j-A_1x_1^3
\end{equation}
and for the high frequency ones
\begin{equation}
  \label{rq7}
 M_j\ddot x_j=-k_jx_j+\sum_{i\ne j}I_{ji}x_i-A_jx_j^3, 
\qquad j\ne 1.
\end{equation}
For a slow motion one can set the acceleration term $M_j\ddot x_j = 0$ 
in Eq.~(\ref{rq7}). For $I\ll
M\omega_0^2$, we have $x_j\ll x_1$ (see below). Therefore we can
neglect also the anharmonicity force term $-A_jx_j^3$ and the interaction
terms $I_{ji}x_i$ ($i\ne 1$) between the high frequency oscillators and
get from Eq.~(\ref{rq7})
\begin{equation}
  \label{kl1}
  x_j = (I_{j1}/k_j)x_1, \qquad j\ne 1 .
\end{equation}
According to Eq.~(\ref{tb2}) we see that
$x_j\simeq(I/M\omega_0^2)x_1\ll x_1$.
Inserting this value of $x_j$ into Eq.~(\ref{ss3}) we finally get
a reduced equation of motion of the low-frequency oscillator
\begin{equation}
  \label{tk4}
 M_1\ddot x_1 = -(k_1-\kappa)x_1-A_1x_1^3=
-\frac{dU_{\rm eff}(x_1)}{dx_1}   
\end{equation}
where
\begin{equation}
  \label{al6}
  \kappa = \sum_{j\ne 1} \frac{I_{1j}^2}{k_j} \simeq \frac{I^2}{M\omega_0^2} 
\end{equation}
and the effective potential energy for the slow motion is
\begin{equation}
U_{\rm eff}(x_1)=\frac{1}{2}(k_1-\kappa)x_1^2 + \frac{1}{4}A_1x_1^4 .
\label{1al}
\end{equation}

The physical origin for this reduction to a one-mode
approximation is the {\em adiabatic approximation} where the high
frequency modes adiabatically follow the slow low-frequency 
motion~\cite{sethna81}.
As a result the interaction between the low and high frequency
oscillators renormalizes the quasielastic constant $k_1$ for the low
frequency motion to the new effective value
\begin{equation}
  \label{fs3}
  k=k_1-\kappa .
\end{equation}

For $k_1>\kappa$ the potential (\ref{1al}) is a one-well potential
and the cluster of oscillators is stable, the equilibrium
positions of all oscillators $x_i=0$.
For $k_1<\kappa$ the renormalized quasielastic constant $k$
is negative and the cluster is unstable. The effective potential
(\ref{1al}) in this case is a symmetric double-well potential. This
is what we call the {\em vibrational instability}. As a result of the
instability the low-frequency oscillator is displaced to one of the
two new minima
\begin{equation}
  \label{ov3}
  x_{10}=\pm\sqrt{(\kappa-k_1)/A_1}=\pm\sqrt{|k|/A_1} ,
\end{equation}
while the displacements of the high frequency ones are
$x_{j0}=(I_{1j}/k_j)x_{10}\ll x_{10}$ ($j\ne 1$) and are much smaller.
The barrier height between the minima is
\begin{equation}
  \label{pl6}
  V=\frac{(\kappa-k_1)^2}{4A_1} = \frac{k^2}{4A_1} .
\end{equation}

As follows from Eq.~(\ref{1al}) the new lowest frequency of the
system of $s$ coupled oscillators  is given by
\begin{equation}
\omega^2 =\left\{{\displaystyle
(k_1-\kappa)/M_1 = k/M_1, \quad \kappa<k_1\, ,\atop
\displaystyle 2(\kappa-k_1)/M_1 = 2|k|/M_1,\quad
\kappa>k_1\,.}\right.
\label{1}
\end{equation}
The first case $\kappa<k_1$ corresponds to a vibration in the
minimum of a one-well potential (\ref{1al}) while the second case
$\kappa > k_1$ corresponds to a vibration in either of the two wells
of a double-well potential (\ref{1al}). It is remarkable that
for weak interaction $I\ll M\omega_0^2$ 
the strength of the anharmonicity $A_1$
does not enter the renormalized frequency (\ref{1}).

Using the Holtsmark method~\cite{H}, we derived in our previous
paper~\cite{GPS} the normalized distribution function of $\kappa$
\begin{equation}
\rho(\kappa)=\frac{1}{\sqrt{2\pi}}\frac{B}{\kappa^{3/2}}\exp
\left(-\frac{B^2}{2\kappa} \right)
\label{dist1}
\end{equation}
where
\begin{equation}
B=\frac{\pi}{3}\sqrt{\frac{\pi}{2}}\,\frac{Jn_0}{\sqrt{M}}
\left\langle\frac{1}{\omega} \right\rangle_0 \equiv
\omega_c\sqrt{M} .
\label{B1}
\end{equation}
Here $Jn_0\simeq I$, and $\left\langle 1/\omega\right\rangle_0\simeq
1/\omega_0$ is the $\omega^{-1}$ moment of the normalized initial
DOS, $g_0(\omega)$. This formula can serve as a definition of the
{\em important characteristic quantities} 
\begin{equation}
  \label{lc8}
 \omega_c\simeq I/M\omega_0 ,\qquad \mbox{and}
\qquad k_c\equiv M\omega_c^2=B^2.
\end{equation}
The physical meaning of these quantities is that the typical clusters with
frequencies $\omega_1\lesssim\omega_c$ become unstable due to the
interaction between the soft oscillator and the surrounding high
frequency ones. Thus the characteristic frequency $\omega_c$ indicates the onset of 
the {\em mechanical instability region}. We will see below that the creation 
of TLS's and the formation of the Boson peak occur in this region. 

\begin{figure}[htb]
\includegraphics[bb=40 110 430 430,totalheight=8cm,keepaspectratio]
{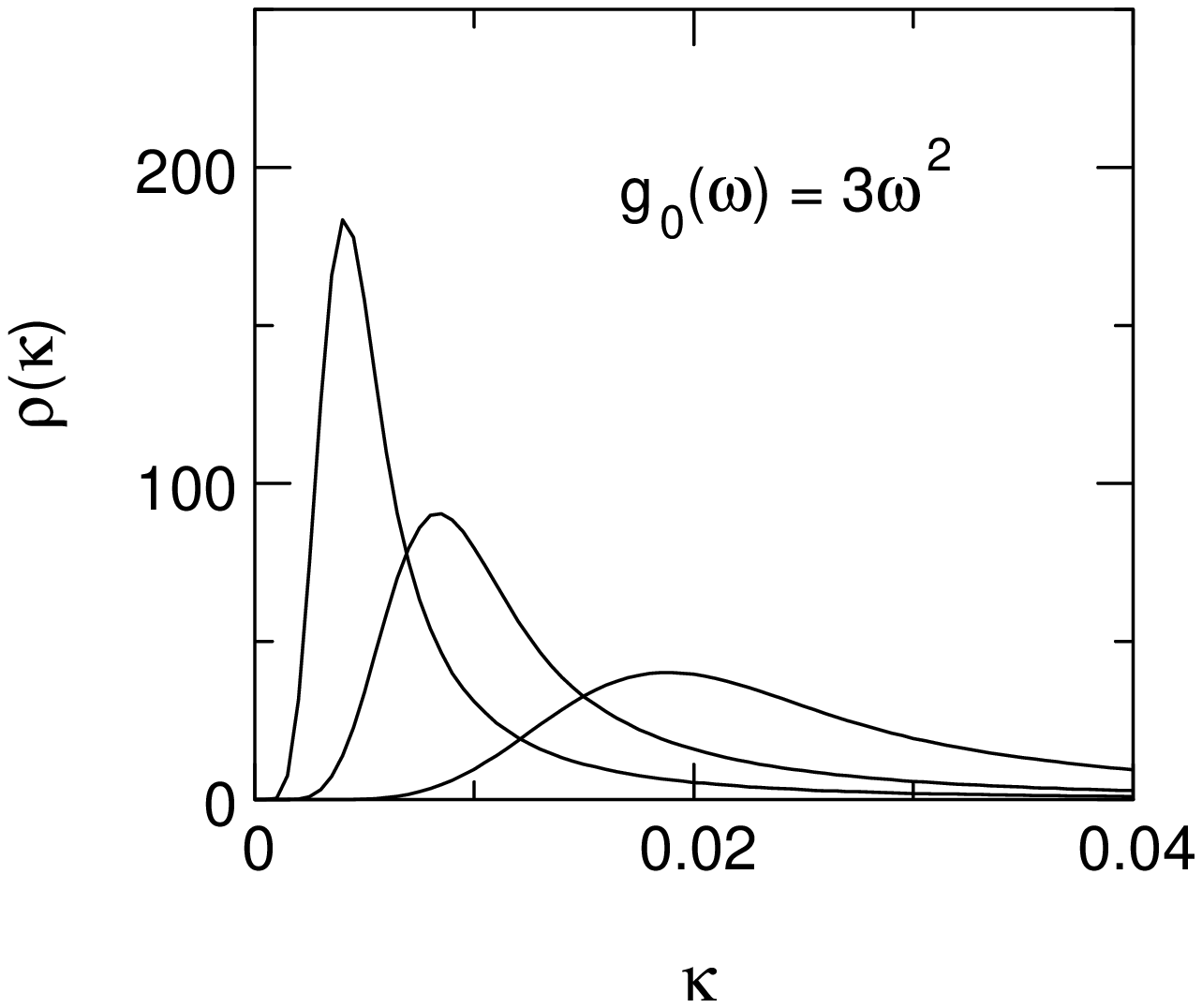}
\caption{Distribution function $\rho(\kappa$), calculated as ensemble
average by exact
diagonalization of systems of $N=2197$ oscillators with $g_0(\omega) =
3\omega^2$ and $J=0.07, 0.10$ and $0.15$ (from left to right).}
\label{fig_rhok}
\end{figure}

However, in the present paper this particular form of the function
$\rho(\kappa)$ is not suitable. The reason is the long-range power tail of
this function, $\rho(\kappa)\propto 1/\kappa^{3/2}$ for $\kappa\gg
k_c$. This tail leads to divergent integrals for large $\kappa$-values when
calculating averages of the type $\left<\kappa^\nu\right>$ for
$\nu\geq 1/2$. As follows from Eqs.~(\ref{al6}) and (\ref{hto1})
the long-range tail of the distribution is related to close pairs with
small distances between the low and high frequency oscillators,
$r_{ij}\ll n_0^{-1/3}$. However, usually the distance between the
HO's in a glass can not be arbitrarily small and, therefore,  
the function $\rho(\kappa)$ drops faster and approaches zero as
$\kappa^{-(n+3)/2}$ (for $g_0(\omega)\propto\omega^n$, with $n>0$). In the
following we do not need the precise analytical form of this
function. It will be sufficient to know that this function decays
sufficiently rapidly for small and large $\kappa$ 
with a characteristic scale $\kappa\simeq k_c\simeq
I^2/k_0$. In Fig.~\ref{fig_rhok} this function is shown for
different interaction strengths $J$ ($n_0=M=\omega_0=1$, see 
Ref.~\cite{GPS} for details). 
%%%%%%%%% Picture 1   %%%%%%%%%%%%%%%%
%%%%%%%%% End Picture 1 %%%%%%%%%%%
Knowing the function $\rho(\kappa)$, we can calculate the distribution
function of the renormalized quasielastic constants, $\Phi(k)$. Let
$F(k_1)$ be a normalized distribution function of quasielastic
constants $k_i$ in Eq.~(\ref{potgen}). In the case of equal masses of the oscillators,
$M_i=M$ it is related to the normalized initial DOS $g_0(\omega)$ as
follows
\begin{equation}
  \label{qw1}
  F(k)=\frac{g_0(\omega)}{2M\omega} , \qquad \mbox{where} \qquad k=M\omega^2 
\end{equation}
and the normalized distribution function $\Phi(k)$ is given as
\begin{equation}
  \label{im3}
  \Phi(k)=\left<\delta(k-k_1+\kappa) \right>_{k_1,\kappa} =
\int\limits_0^\infty dk_1 F(k_1)\int\limits_0^\infty d\kappa \rho(\kappa)
\delta(k-k_1+\kappa) .
\end{equation}

Integrating over the delta-function, it is convenient to present the
expression for $\Phi(k)$ for positive and negative $k$
separately. We have from Eq.~(\ref{im3})
\begin{equation}
\Phi(k)=\int\limits_0^\infty dk_1 F(k_1)\rho(k_1+|k|), \quad \mbox{for}
\quad k<0 ,
\label{ty2}
\end{equation}
and
\begin{equation}
\label{bx1}
\Phi(k)=\int\limits_0^\infty d\kappa\rho(\kappa)F(k+\kappa) ,
\quad\mbox{for}\quad k>0.
\end{equation}

\begin{figure}[htb]
\includegraphics[bb=40 110 430 450,totalheight=8cm,keepaspectratio]
{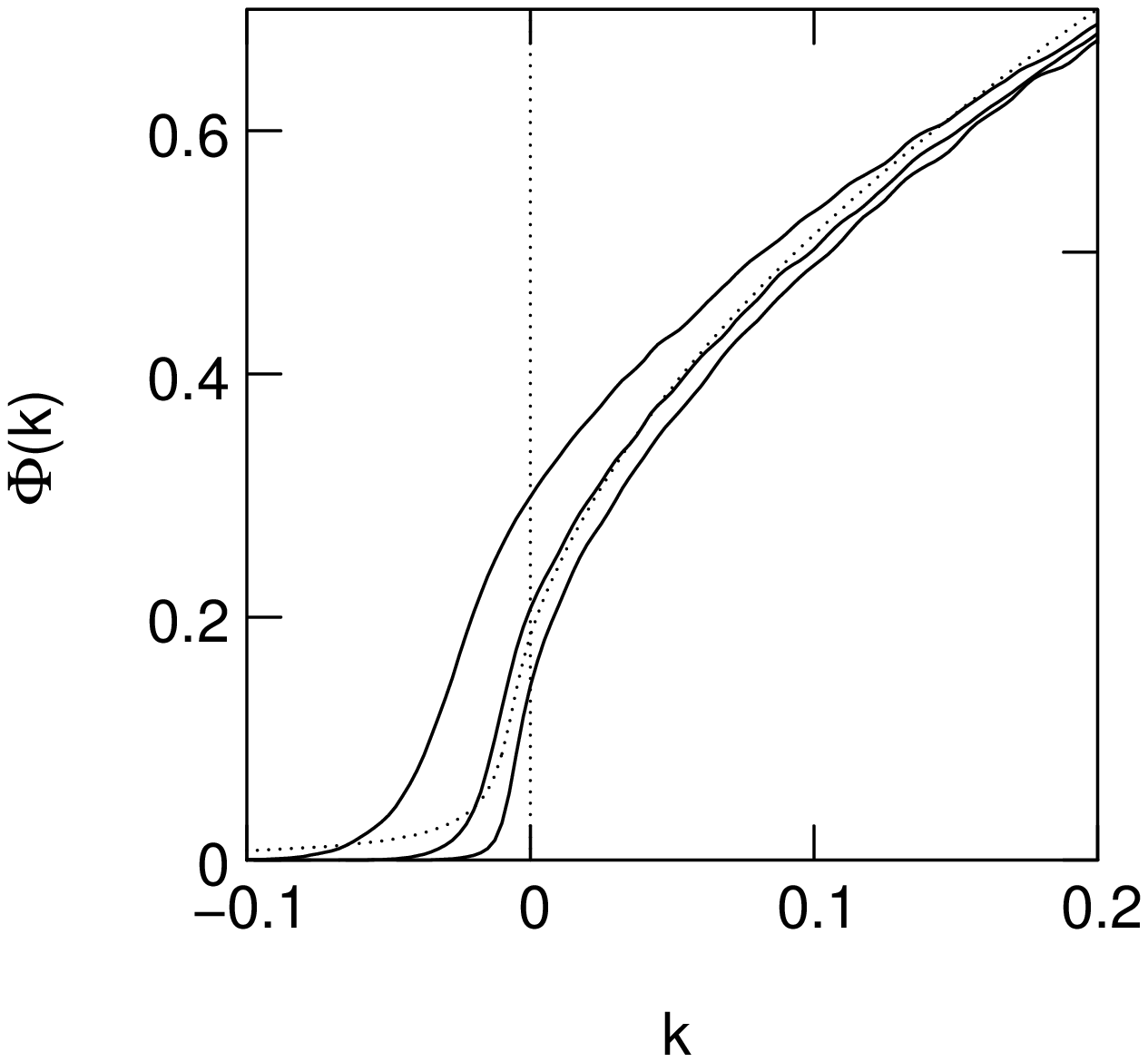}
\caption{Distribution function of the
renormalized quasielastic constants, $\Phi(k)$, calculated as ensemble average by exact
diagonalization of a systems of $N=2197$ oscillators with $g_0(\omega)
= 3\omega^2$ ($F(k) = 3\sqrt{k}/2$) and $J=0.07,0.10,0.15$ (solid curves, from right to
left). Dotted curve: Result of convolution, Eq.~(\ref{im3}), for
$J=0.1$.}
\label{fig_G_unrel}
\end{figure}

Since the distribution function $\rho(\kappa)$ is nonvanishing only for
$\kappa\lesssim k_c$ and rapidly drops to zero for $\kappa
\gtrsim k_c$ we conclude from Eqs.~(\ref{ty2}) and (\ref{bx1}) that
the function $\Phi(k)$ for $|k|\ll k_c$ is approximately a constant
\begin{equation}
  \label{un4}
 \Phi(k)\approx \Phi(0)=
\int\limits_0^\infty d\kappa\rho(\kappa)F(\kappa)\approx F(k_c), \qquad
\mbox{for}\qquad |k|\ll k_c .
\end{equation}
For negative $k$ and $|k|\gtrsim k_c$ the function $\Phi(k)$
rapidly drops. For positive $k\gtrsim k_c$, $\Phi(k)\approx F(k)$.
In Fig.~\ref{fig_G_unrel} this function is shown for
different interaction strength $J$ ($n_0=M=\omega_0=1$, see 
Ref.~\cite{GPS} for details).
%%%%%%%%% Picture 2   %%%%%%%%%%%%%%%%
%%%%%%%%% End Picture 2     %%%%%%%%%%%

\section{interaction between low-frequency oscillators}
\label{intlf}

In the previous section we have considered the effect of the
interaction between a low-frequency oscillator and the surrounding high
frequency oscillators in a cluster. As a result of this interaction
the quasielastic constant $k_1$ has been renormalized to a new
effective value $k=k_1-\kappa$. Negative values of $k$ indicate a
vibrational instability of the cluster. So far we have neglected the
interaction between low-frequency oscillators belonging to different
clusters. This interaction is much weaker and cannot produce a new
instability. However, it causes internal random static forces acting
on the low-frequency oscillators. As we have seen in the previous
section, in the case of instability the interaction between low
frequency and high frequency oscillators shifts the positions of
their minima (static displacements). These shifts in turn act as
forces if we take the interaction between unstable low-frequency HO's
into account. As was shown in our previous paper~\cite{GPS} (see also 
Section~\ref{bospeak}) these forces are responsible for the universal 
$g(\omega) \propto \omega^4$ (see also Ref.~\cite{GC}) 
dependence of the excess vibrational density of states for low 
frequencies.

One can insert these forces into Eq.~(\ref{1al}) as
a linear term $-fx_1$ where $f$ is the internal random force
created by the other unstable low-frequency oscillators.
The effective potential energy then reads
\begin{equation}
U_{\rm eff}(x)=-fx+\frac{1}{2}kx^2+\frac{1}{4}Ax^4 .
\label{t1}
\end{equation}
Here and henceforth the index 1 will be omitted.

The distribution function of the random forces $P(f)$ has been
obtained in our previous paper~\cite{GPS}
\begin{equation}
P(f)=\frac{1}{\pi}\frac{\delta f}{f^2+(\delta f)^2} ,
\label{t3}
\end{equation}
where $\delta f$ is the width of the distribution. The Lorentzian
form of the distribution is related to the fact that the forces
between harmonic oscillators decay as $r_{ij}^{-3}$ [see
Eq.~(\ref{hto1})].

One can estimate the width, $\delta f$, of the distribution
as follows. The static force $f_i$ exerted on the $i$th oscillator by
the $j$th one is
\begin{equation}
f_i=I_{ij}x_{j0}.
\label{f}
\end{equation}
Its characteristic value is given by
displaced harmonic oscillators with frequencies of order of
$\omega_c$. For these oscillators we have
\begin{equation}
I_{ij}^{(c)}\approx Jn_c, \quad J\approx I/n_0\approx
M\omega_c\omega_0/n_0\approx M\omega_c/n_0g_0(\omega_0).
\label{int2}
\end{equation}
Here $n_c\approx n_0g_0(\omega_c)\omega_c$ is the concentration of
these unstable harmonic oscillators (double-well potentials) while
$n_0$ is the total concentration of HO's. Due to the normalization
condition $\omega_0g(\omega_0)\approx 1$. From Eq.~(\ref{ov3}) it
follows that the characteristic static displacement of these
unstable oscillators is $x_{j0}\approx
\sqrt{k_c/A}=\omega_c\sqrt{M/A}$. As a result, one gets from
Eq.~(\ref{f}) (see Ref.~\cite{GPS})
\begin{equation}
\delta f \approx Jn_c\, \omega_c\sqrt{\frac{M}{A}}\approx
M \sqrt{\frac{M}{A}}\,\omega_c^3\,
\frac{g_0(\omega_c)}{g_0(\omega_0)} .
\label{t4}
\end{equation}

As mentioned already in the beginning of this section, these
internal random forces do not produce a new vibrational instability.
However, they  can transform some double-well potentials into
single-well ones. For $k>0$ the potential (\ref{t1}) is always
one-well whereas for $k<0$ the potential (\ref{t1}) is
double-well for small forces $|f|<f^\star_k$ where
\begin{equation}
\label{ue7}
f^\star_k=2|k|^{3/2}/3\sqrt{3A} .
\end{equation}
For $|f|>f^\star_k$ (and $k<0$) (\ref{t1}) is a one-well
potential. For $|f|=f^\star_k$ ($k<0$) the potential (\ref{t1}) is
one-well with a bending point.

It is interesting to compare the width of the distribution $\delta f$
with the characteristic value of $f^\star_k$. They become equal for
$|k|=k^\star\equiv M\left(\omega^\star\right)^2$ where
\begin{equation}
  \label{yc3}
  \omega^\star \approx \omega_c \left[g_0(\omega_c)/g_0(\omega_0)
 \right]^{1/3}\ll \omega_c .
\end{equation}
The strong inequality $\omega^*\ll \omega_c$ occurs if both
$\omega_c\ll\omega_0$ and $g_0(\omega_c)\ll g_0(\omega_0)$. 
We shall see in Section~\ref{bospeak} that the frequency
$\omega^\star$ plays a role of the Boson peak frequency.

\section{Two-level systems}
\label{TLS}

As follows from Eq.~(\ref{t1}) negative values of $k$ for $f=0$
correspond to symmetric double-well potentials. In a purely classical
treatment the oscillator will vibrate in either of the wells.  Taking
quantum mechanics into account, there will be a finite probability of
penetration through the barrier separating the two wells, i. e.
there is a finite tunneling probability. This causes a splitting of
the vibrational levels. We are interested in the lowest pair of
levels. This constitutes a two-level system (TLS). These systems are
ubiquitous in glasses and determine their low temperature
properties~\cite{phillips:81}.

Tunneling systems can be described effectively in terms of a tunnel
splitting $\Delta_0$ and an asymmetry $\Delta$. We
will derive expressions for these quantities and their distributions.
Neglecting the linear force term in Eq.~(\ref{t1}), the tunnel splitting
is given in the WKB approximation as
\begin{equation}
  \label{de1}
  \Delta_0\approx W\exp\left(-S/\hbar \right), \quad
S=\int\limits_{-x_0}^{x_0}|p|dx =2\int\limits_0^{x_0}
\sqrt{2M\left[U_{\rm eff}(x)+V \right]}dx .
\end{equation}
Here we approximated the dependence of the prefactor on the
vibrational frequency by an order of magnitude estimate $W$ 
(see also Ref.~\cite{DPR})
\begin{equation}
\label{wdef}
W=\frac{\hbar}{2}\left(\frac{\hbar A}{2 M^2}\right)^{1/3}
\end{equation}
that is of the order of the interlevel spacing in a purely quartic
potential $V(x)=Ax^4/4$. From experiment, using the the soft-potential
model~\cite{KKI,DPR,BGGS,RB,BGGPRS,SP}, one finds values  $W$ for different
glasses of the order of a few Kelvin, e. g. for vitreous silica
$W\approx 4$~K. The positions of the minima in the symmetric
double-well potential $U_{\rm eff}(x)$ (for $f=0$) are denoted by
$\pm x_0=\pm\sqrt{|k|/A}$ and $V=k^2/4A$ is the barrier hight [see
Eqs.~(\ref{ov3}) and (\ref{pl6})].

Evaluating the integral in Eq.~(\ref{de1}) we get
\begin{equation}
  \label{de2}
  S=\frac{2\sqrt{2}}{3}\frac{|k|^{3/2}M^{1/2}}{A}
\end{equation}
and
\begin{equation}
\label{t5}
\Delta_0=W\exp\left(-\frac{2\sqrt{2}}{3}\frac{|k|^{3/2}M^{1/2}}
{\hbar A} \right)=W\exp\left(-\frac{\sqrt{2}}{24}
\frac{\hbar^3|k|^{3/2}}{M^{3/2}W^3} \right).
\end{equation}

The second quantity characterizing the TLS is the asymmetry of the
two-well configuration. For $|f|\ll f^\star_k$ we have from
Eq.~(\ref{t1})
\begin{equation}
\label{t6}
\Delta=2fx_0=2f\sqrt{|k|/A}.
\end{equation}

We are interested in the two quantum states with the lowest energies.
These states belong to both wells. TLS are often described by
the interlevel distance $E$ and the dimensionless  tunneling parameter $p$:
\begin{equation}
\label{t7}
E=\sqrt{\Delta_0^2+\Delta^2},  \qquad
p=\left(\Delta_0/E\right)^2.
\end{equation}

The Jacobian of the transformation from the variables $|k|$ and $f$
to $E$ and $p$ is
\begin{equation}
\label{jac1}
{\cal J}=\frac{\partial(|k|,f)}{\partial(E,p)}=
\left(\frac{2}{9} \right)^{1/3}2^{5/2}
\frac{M^{3/2}W^{5/2}}{\hbar^3} \frac{L^{-2/3}}{p\sqrt{1-p}}, \qquad
{\mbox{ where}} \qquad L=\ln \frac{W}{E\sqrt{p}} .
\end{equation}
In the new variables the distribution function reads
\begin{equation}
\label{t8}
F(E,p)=n_0P(0)\Phi(0)|{\cal J}| =
\left(\frac{2}{9} \right)^{1/3}2^{5/2}n_0P(0)\Phi(0)
\frac{M^{3/2}W^{5/2}}{\hbar^3} \frac{L^{-2/3}}{p\sqrt{1-p}}.
\end{equation}
Here we have replaced  $P(f)$ and $\Phi(k)$ by $P(0)$ and $\Phi(0)$.
This can be justified by estimating the relevant ranges of $f$ and $|k|$
given by Eq.~(\ref{cond1}). Using Eqs.~(\ref{t6}), (\ref{t5})
and (\ref{t4}) we express $f$ in terms of $\Delta_0$ and $\Delta$
\begin{equation}
\label{d14}
\frac{f}{\delta f}\approx 4\left(\frac{2}{9}\right)^{1/6}L^{-1/3}\,
\frac{E\sqrt{1-p}}{W} \left(\frac{W}{\hbar\omega_c}\right)^3
\frac{g_0(\omega_0)}{g_0(\omega_c)} .
\end{equation}
Taking rough estimates $L=10$, $E=1\,K$, $W=4\,K$,
$\hbar\omega_c=100\,K$, $\omega_0/\omega_c=3$, and
$g_0(\omega)\propto\omega^2$ one finds a typical value
$f/\delta f\approx 2\cdot 10^{-4}$. To estimate $|k|$ we derive
from Eq.~(\ref{t5})
\begin{equation}
\label{d15}
|k|/k_c = 4\left(9/2\right)^{1/3}L^{2/3}
\left(W/\hbar\omega_c\right)^2  .
\end{equation}
This set of parameters gives the typical value of
$|k|/k_c\approx 0.05$. Therefore, in the range of parameters where
the notion of TLS's is applicable, the characteristic values of
$f$ and $|k|$ satisfy the conditions
\begin{equation}
  \label{cond1}
 f\ll\delta f, \quad \mbox{and} \quad |k|\ll k_c .
\end{equation}

Our result can be compared with the standard tunneling
model~\cite{phillips:81} where the distribution functions are
\begin{equation}
\label{t9}
P(\Delta, \Delta_0)=\frac{\overline{P}}{\Delta_0},
\qquad F(E,p)=\frac{1}{2}
\frac{\overline{P}}{p\sqrt{1-p}} 
\end{equation}
with a constant density of tunneling states $\overline{P}$.
Comparing these distributions with Eq.~(\ref{t8}),
one gets
\begin{equation}
\label{t10}
\overline{P}=\left(\frac{2}{9} \right)^{1/3}2^{7/2}n_0P(0)\Phi(0)
\frac{M^{3/2}W^{5/2}}{\hbar^3} L^{-2/3} .
\end{equation}
Both distributions, Eqs.~(\ref{t8}) and (\ref{t9}),
coincide regarding their dependencies on $E$ and $p$, apart
from the factor $L^{-2/3}$ describing a weak logarithmic dependence on
$E$ and $p$. The same factor is found in the
soft potential model (see  Refs.~\cite{KKI,IKP}).

To compare the tunneling strengths of the TLS's with experiment we
study their interaction with strain, described by the deformation
potential $\gamma$. According to Eq.~(\ref{int1}), the variation
of asymmetry $\Delta_{\varepsilon}$ due to a strain $\varepsilon$ is
\begin{equation}
\label{d1}
\Delta_{\varepsilon}=2\Lambda x_0\,\varepsilon =
2\Lambda \varepsilon\sqrt{|k|/A} .
\end{equation}
The deformation potential is defined as
\begin{equation}
\label{d2}
\gamma=\frac{1}{2}\frac{\partial\Delta_{\varepsilon}}{\partial
\varepsilon}
\end{equation}
and from Eqs.~(\ref{d1}), (\ref{t5}) and (\ref{wdef}) follows
\begin{equation}
\label{d3}
\gamma=\Lambda\sqrt{\frac{|k|}{A}}=\frac{3^{1/3}}{2^{7/6}}
\frac{\hbar\Lambda}{\sqrt{M W}} L^{1/3}.
\end{equation}

In the standard tunneling model the TLS's are often characterized
by the  dimensionless tunneling strength $C$, given by Eq.~(\ref{yu6g}).
For different glasses its value varies between 10$^{-3}$ and 10$^{-4}$.
Using  Eqs.~(\ref{t10}) and (\ref{d3}) one gets
\begin{equation}
\label{d5}
C=2\sqrt{2}n_0P(0)\Phi(0)\frac{\Lambda^2}{\rho v^2}
\frac{W^{3/2}\sqrt{M}}{\hbar} .
\end{equation}
This value is independent of $E$ and $p$ as in the standard tunneling model.

The different factors entering the expression for $C$ can be estimated
from our model as
\begin{equation}
\label{d6}
\Lambda^2/\rho v^2=J\approx I/n_0, \qquad I\approx M\omega_c\omega_0 ,  \qquad
\Phi(0)\approx F(k_c)=g_0(\omega_c)/2M\omega_c
\end{equation}
and from Eqs.~(\ref{t3}), (\ref{t4}) and (\ref{wdef})
\begin{equation}
\label{d7}
P(0)=\frac{1}{\pi\delta f}\approx \frac{4}{\pi} \,\frac{W^{3/2}}
{\hbar^2\sqrt{M}}\,\frac{g_0(\omega_0)}{\omega_c^3g_0(\omega_c)} .
\end{equation}
As a result we arrive at the important estimate
\begin{equation}
\label{d8}
C\approx \frac{4\sqrt{2}}{\pi}\left(\frac{W}{\hbar\omega_c} \right)^3
\end{equation}
that is independent of the initial DOS of HO's $g_0(\omega)$ and in
this sense is {\em universal}. It only depends on the characteristic energy
$W$ and the frequency $\omega_c$ that is proportional to the
interaction $I$. The larger the interaction between the original
oscillators the smaller is constant $C$, $C \propto I^{-3}$. 

In the discussion of the Boson peak (Section V) we will see that $\omega_c$
is two or three times larger than the Boson peak frequency $\omega_b$
(or $\omega^\star$) which slightly depends on initial DOS $g_0(\omega_c)$ 
(see Eqs.~(\ref{fre1}), (\ref{k}) and Fig.~\ref{fig:bpeak}). 
Using values appropriate to SiO$_2$, $W=4$~K and
$\hbar\omega_c = 100$~K we get
\begin{equation}
\label{estimate_C}
C \approx 10^{-4}.
\end{equation}
Thus the unified approach of this paper gives a value of the
tunneling strength $C$ in good agreement with experiment.
Since values of $C$ and $W$ are well known from experiment for many 
glasses~\cite{DPR}, Eq.~(\ref{d8}) can be used to estimate the important 
characteristic energy $\hbar\omega_c$ giving onset of the vibrational 
instability in glasses
\begin{equation}
\hbar\omega_c \approx W C^{-1/3} .
\label{st6r}
\end{equation}
It was demonstrated in Ref.~\cite{ParLaer} that 
taking into account the experimental data for $C$ and $W$ 
this energy is indeed correlated with position of the Boson peak in 
glasses,  $\omega_b\simeq\omega_c$.

The two factors entering $C$, Eq.~(\ref{yu6g}), can be estimated
separately as 
\begin{equation}
\label{d9}
\overline{P}\approx\frac{n_0}{\hbar\omega_0}
\left(\frac{W}{\hbar\omega_c}\right)^4 L^{-2/3},
\qquad \frac{\gamma^2}{\rho v^2}\approx\frac{\hbar\omega_0}{n_0}\,
\frac{\hbar\omega_c}{W}\, L^{2/3} .
\end{equation}
The first of these quantities is $\propto \hbar^{1/3}$ and the second
$\propto \hbar^{2/3}$ and thereby $C \propto \hbar$.  Since $C\ll 1$ is a
dimensionless quantity it can be represented in the form
$C=\hbar/\widetilde{S}$ where $\widetilde{S}\gg \hbar$ is some classical action. From
Eqs.~(\ref{d8}), (\ref{de2}) and (\ref{wdef}) we get
\begin{equation}
\label{tr3}
C\approx\frac{1}{3\pi}\frac{\hbar}{S_c}\ll 1, \quad
S_c=\frac{2\sqrt{2}}{3}\frac{|k_c|^{3/2}M^{1/2}}{A} ,
\end{equation}
i.e. $\widetilde{S}\approx 3\pi S_c$. The classical action $S_c$ corresponds
to a typical double-well potential (\ref{t1}) with $|k|=k_c$ (and $f=0$). 

Using the estimates $\omega_c/\omega_0 \approx 1/3$, $W=4$~K and
$\hbar\omega_c = 100$~K we estimate
the concentration of tunneling systems with energies in the range
$0<E<W$ and tunneling parameter $p \simeq 1$ as
\begin {equation}
\label{eq_ntls}
n_{\rm TLS} \simeq \overline{P}W \simeq n_0\frac{\omega_c}{\omega_0}
\left(\frac{W}{\hbar\omega_c}\right)^5 L^{-2/3} \approx 3\cdot10^{-8}n_0.
\end{equation}
The number of active TLS's is thus less than {\em one for ten million of
oscillators}. This explains why the concentration of observed TLS's
in glasses is so small. Since according to Eq.~(\ref{eq_ntls}) $n_{\rm
  TLS}\propto I^{-4}$ the number of TLS's decreases rapidly with increasing
interaction strength $I$. 

It would be instructive to derive a dimensionless parameter $C_{\rm cl}$ 
by a {\em classical} procedure (neglecting the tunneling
probability $\Delta_0$). We take the width of the force
distribution from Eq.~(\ref{t4}) and estimate the typical
asymmetry $\Delta_c$ from Eq.~(\ref{t6})
\begin{equation}
\label{d10}
\delta f\approx I^{(c)} x_0^{(c)}\approx Jn_c\sqrt{k_c/A}, \qquad
\Delta_c\approx\delta f\sqrt{k_c/A}\approx Jn_c\cdot(k_c/A) .
\end{equation}
With $n_c$ the concentration of double-well potentials
we get the classical estimate for their density of states
\begin{equation}
\label{d11}
\overline{P}_{c}\approx\frac{n_c}{\Delta_c}\approx\frac{A}{Jk_c}
\approx \frac{n_0}{\hbar\omega_0}\left(\frac{W}{\hbar\omega_c}\right)^3
\end{equation}
that is independent of $\hbar$.

For the deformation potential $\gamma_c$ we have from
Eqs.~(\ref{d1}) and (\ref{d2}) the estimate
\begin{equation}
\label{d12}
\gamma_c\approx\Lambda\sqrt{k_c/A},\qquad\mbox{and} \qquad
\frac{\gamma_c^2}{\rho v^2}\approx J\frac{k_c}{A}\approx
\frac{\hbar\omega_0}{n_0}\left(\frac{\hbar\omega_c}{W} \right)^3,
\end{equation}
also independent of $\hbar$. Finally
\begin{equation}
\label{d13}
C_{\rm cl}=\frac{\overline{P}_c\gamma_c^2}{\rho v^2} \approx 1 ,
\end{equation}
i.e. the dimensionless parameter $C_{\rm cl}$ in this
classical approach is of the
order of unity which is a consequence of the $1/r^3$
interaction between the TLS's~\cite{Klein,YuLeg,CCYu,Coop}.
We wish to emphasize that $C_{\rm cl}$ (unlike $C$) does not
determine any physical property of glasses.

The reason for the difference between the two approaches (quantum
and classical) is the following. In the classical approach we take
all the double-well potentials into account. They have typically
$|k|\simeq k_c$. Their concentration, $n_c$, is unimportant
since it is canceled in Eq.~(\ref{d11}) for
$\overline{P}_c$. In the quantum approach only the
small portion of TLS's which are able to tunnel
(they have $|k|<<k_c$) contribute to the observable quantities.
For all other TLS's the high barriers $V$ and asymmetries $\Delta
\sim \Delta_c$ prevent tunneling (but they contribute to the internal
random static force $\delta f$).

To further clarify this point let us consider TLS's with a quasielastic
constant $k$ in the interval $\Delta k\simeq k$ where $|k|\ll
k^\star \ll k_c$ [see Eq.~(\ref{yc3})]. Their concentration $n_k$,
asymmetry $\Delta_k$ and the deformation potential $\gamma_k$
are given by
\begin{equation}
  \label{ni6}
  n_k\approx n_0|k|\Phi(0)(f^\star_k/\delta f), \qquad
\Delta_k\approx f^\star_k\sqrt{|k|/A}, \qquad
\gamma_k=\Lambda\sqrt{|k|/A} .
\end{equation}
In this expression for $n_k$ we took into account, that for
$|k|\ll k^\star$, only the small fraction of all potentials, where
$f^\star_k/\delta f\ll 1$, is of double-well type. Keeping in mind that
$\Phi(0)\approx F(k_c)$ we get  the density of states
\begin{equation}
  \label{ak1}
  \overline{P}_k \approx \frac{n_k}{\Delta_k}\approx
\frac{n_0|k|F(k_c)}{\delta f\sqrt{|k|/A}}
\end{equation}
and the parameter $C_k$
\begin{equation}
  \label{me1}
  C_k = \frac{\overline{P}_k\gamma_k^2}{\rho v^2}\approx
\frac{n_0F(k_c)J|k|^{3/2}}{\delta f\sqrt{A}}\approx
\left(\frac{|k|}{k_c} \right)^{3/2}\ll 1.
\end{equation}
Here we have used Eq.~(\ref{d10}) for $\delta f$ and the estimate
$n_c\simeq n_0k_cF(k_c)\simeq n_0\omega_c g_0(\omega_c)$.

If we now fix $|k|$ by the condition that the exponent in
Eq.~(\ref{t5}) is of order of unity, i.e.  $|k|\simeq MW^2/\hbar^2$,
we reproduce our quantum result, Eq.~(\ref{d8}), $C_k\approx
(W/\hbar\omega_c)^3$. The classical result, $C_{\rm cl}$, 
Eq.~(\ref{d13}) would be recovered for $k\approx k_c$ when
$C_k\approx 1$.  We conclude that {\it the physical reason for
smallness of the parameter} $C$ {\it for TLS's in glasses is the
scarcity of those TLS's that are able to tunnel compared to their
total number}.

Again we can compare our results with the standard tunneling
model~\cite{phillips:81}. In this model the tunneling amplitude
$\Delta_0=\hbar\omega_0\exp(-\lambda)$ and the dimensionless
parameter $\lambda$ is uniformly distributed in the interval
$\lambda_{\rm min} < \lambda<\lambda_{\rm max}$. The lowest
value $\lambda_{\rm min}\simeq 1$. According to Eq.~(\ref{de1})
$\lambda=S/\hbar$, and therefore the maximal value $\lambda_{\rm
max}\approx S_c/ \hbar$. Taking into account Eq.~(\ref{tr3}) we get
\begin{equation}
  \label{pz3}
  \lambda_{\rm max}\approx S_c/\hbar\approx 1/3\pi C .
\end{equation}
Thus $\lambda_{\rm max}$ is related to the small parameter $C$.
For SiO$_2$, $C=3\cdot 10^{-4}$ and $\lambda_{\rm max}\approx 350$.

\section{the boson peak}
\label{bospeak}

In this section we relate the results obtained for the TLS's
to the Boson peak properties --- see Ref.~\cite{GPS}. For
this we calculate the vibrational density of states (DOS)
$g(\omega)$. We start from the case $f=0$ (i.e. neglecting the
interaction between the clusters). There are two types
of harmonic vibrations. In the one-well case, $k>0$,
according to Eq.~(\ref{un4}) the distribution function of $k$ for
$k\ll k_c$ is constant, $\Phi(k)\approx \Phi(0)$. Therefore, since
according to Eq.~(\ref{1}) (top) $k=M\omega^2$, the renormalized DOS for 
$\omega\ll\omega_c$ is 
\begin{equation}
\label{b2}
\widetilde{g}_{\rm I}(\omega)=2n_0M\Phi(0)\omega .
\end{equation}

For harmonic vibrations in either well of
a symmetric double-well potential [Eq.~(\ref{t1}), $k<0$ and $f=0$]
$|k|=M\omega^2/2$ [see Eq.~(\ref{1}), bottom] and
$\Phi(k)=\Phi(0)$ for $|k|\ll k_c$ ($\omega\ll\omega_c$)
the DOS is
\begin{equation}
  \label{rf3}
\widetilde{g}_{\rm II}(\omega)=n_0M\Phi(0)\omega   .
\end{equation}
It is half of the one-well contribution. The total DOS for $f=0$
and $\omega\ll\omega_c$ is the sum of the two contributions
\begin{equation}
  \label{fv1}
\widetilde{g}_{\rm tot}(\omega) = \widetilde{g}_{\rm I}(\omega) +
\widetilde{g}_{\rm II}(\omega) = 3n_0M\Phi(0)\omega .
\end{equation}
It is a {\em linear} function of $\omega$ independent of the form of the
initial DOS $g_0(\omega)$. This linear behavior follows from the finite
value of $\Phi(0)$.

If the low-frequency HO's were isolated their density of states would
be determined by Eq.~(\ref{fv1}). As we have shown in the
Section~\ref{intlf} there is, however, an interaction between these
oscillators which we have to take into account. According to
Eq.~(\ref{f}) the low-frequency harmonic oscillators, displaced from
their equilibrium positions (and forming the double-well potentials),
create long-range random static forces $f$ acting on other
oscillators. In a purely harmonic case, these linear forces would not
affect the frequencies.  Anharmonicity, however, renormalizes the low
frequency part of the spectrum, a manifestation of the so-called {\em
sea-gull singularity} treated in detail in Ref.~\cite{IKP} 
(see also Ref.~\cite{BGGS}).

We begin with the case  $k>0$. It corresponds to one-well
potentials. Consider an anharmonic oscillator under the action of a
random static force $f$. The effective potential is given by
Eq.~(\ref{t1}) where $\sqrt{k/M}$ is the oscillator frequency in
the harmonic approximation for $f=0$. The force $f$ shifts the
equilibrium position from $x=0$ to $x_0\neq 0$, given by
\begin{equation}
\label{xce8}
Ax_0^3 + kx_0 - f = 0 ,
\end{equation}
where the oscillator has a new (harmonic) frequency
\begin{equation}
\label{kiu6}
M\omega^2_{\rm new} = k+3Ax_0^2 .
\end{equation}
With $\Phi(k)$  as the distribution function of $k$ [see
Eq.~(\ref{bx1})] and $P(f)$ as the distribution of random forces
$f$ [see Eq.~(\ref{t3})] the renormalized DOS is given by
\begin{equation}
\label{des2}
g_{\rm I}(\omega)=n_0\int\limits_0^\infty \Phi(k)dk
\int\limits_{-\infty}^{\infty}dfP(f)\delta\left(\omega -
\omega_{\rm new}\right) .
\end{equation}

Assuming $\omega\ll\omega_c$ and integrating Eq.~(\ref{des2})
with $\Phi(k)=\Phi(0)$ we get the integral
\begin{equation}
  \label{mq2}
g_{\rm I}(\omega) = 2n_0\Phi(0)\frac{M^2\omega^3}{\sqrt{3A}}
\int\limits_0^{M\omega^2}dk\frac{P\left[f(k)\right]}{\sqrt{M\omega^2-k}}
\end{equation}
where according to Eqs.~(\ref{xce8}) and (\ref{kiu6})
\begin{equation}
  \label{de7}
  f(k)=Ax_0^3+kx_0=\frac{1}{3}
\sqrt{\frac{M\omega^2-k}{3A}}(2k+M\omega^2) .
\end{equation}
Taking the Lorentzian distribution, Eq.~(\ref{t3}), for $P(f)$ and
introducing a new variable $t=\sqrt{1-k/M\omega^2}$ we finally get
\begin{equation}
\label{bp1}
g_{\rm I}(\omega) = \frac{12}{\pi}n_0
M\Phi(0) \frac{\omega^2}{\omega^\star}
\left(\frac{\omega}{\omega^\star}\right)^2\int\limits_0^1
\frac{\displaystyle dt}{\displaystyle
1+\left(\omega/\omega^\star\right)^6t^2(3-2t^2)^2}
\end{equation}
with
\begin{equation}
\label{fre1}
\omega^\star = {\sqrt{3}A^{1/6} (\delta f)^{1/3}/\sqrt{M}} .
\end{equation}

The function $g_{\rm I}(\omega)$ depends on a single parameter,
$\omega^\star$ characterizing, as we will see below, the position of the
Boson peak $\omega_b$ ($\omega_b\approx \omega^*$). 
The frequency $\omega^\star$ is determined by the
characteristic value of the random static force $\delta f$ acting on
an HO with the characteristic frequency $\omega_c$. As a result, taking
into account Eq.~(\ref{t4}), we get the estimate
\begin{equation}
\omega^\star \approx \omega_c \left[\frac{g_0(\omega_c)}{g_0(\omega_0)}
\right]^{1/3},\quad \omega^\star \ll \omega_c  .
\label{k}
\end{equation}
Again, as in Eq.~(\ref{1}) in lowest order the anharmonicity $A$ does
not enter this formula. This equation for $\omega^\star$ coincides
with Eq.~(\ref{yc3}) obtained from the condition
$f_k^\star\simeq\delta f$. 

According to Eq.~(\ref{k}), for weak interactions $I$ 
($\omega_c\ll\omega_0$), the frequency of the Boson peak $\omega^*\ll\omega_c$
only in the case when the initial DOS, $g_0(\omega)$, is monotonically (and rapidly)
decreasing (to zero) function of
$\omega$. For example we can take $g_0(\omega)\propto\omega^n$ with $n>0$. Then for
$n=2$ and $\omega_c = \omega_0/3$ we have from
Eq.~(\ref{k}) $\omega^*\approx\omega_c/2$. For the same $n$ and smaller interaction,
$\omega_c=\omega_0/5$, we get $\omega^*\approx\omega_c/3$. In the opposite case if 
the initial DOS drops to zero
too slowly or remains nearly constant, $g_0(\omega)\simeq \mbox{const}$, we have from
Eq.~(\ref{k}) that $\omega^* \simeq \omega_c$. In this case the Boson peak
frequency $\omega^*$ is of the same order as the characteristic frequency 
$\omega_c$.

For small frequencies, $\omega\ll\omega^\star$, only small forces
$f\ll\delta f$ contribute to the integral in Eq.~(\ref{mq2}). In this
case the distribution function $P(f)$ can be approximated by a
constant value, $P(0)$, and we get from Eqs.~(\ref{mq2}) and
(\ref{bp1})
\begin{equation}
  \label{zi7}
g_{\rm I}(\omega) = 4n_0\Phi(0)P(0)\frac{M^{5/2}\omega^4}
{\sqrt{3A}} = \frac{12}{\pi} n_0 M\Phi(0)\omega
\left(\frac{\omega}{\omega^\star}\right)^3\propto\omega^4.
\end{equation}
As a result, at low frequencies the renormalized excess DOS, $g_{\rm I}
(\omega)\propto \omega^4$, Ref.~\cite{IKP,GC}. For sufficiently large
frequencies, $\omega\gg\omega^\star$ (but still smaller than $\omega_c$)
the action of random static forces on the HO spectrum can be
discarded. In this case the integral in Eq.~(\ref{bp1}) is equal to
$(\pi/6)(\omega^\star/\omega)^3$. We recover the linear
DOS, Eq.~(\ref{b2}), $g_{\rm I}(\omega) = 2n_0M\Phi(0)\omega
\propto\omega$.

For $k<0$ the effective potential energy
including a random static force $f$ is given by Eq.~(\ref{t1}). The
simple analysis in Section~\ref{intlf} shows that for sufficiently
small force, $|f|<f^\star_k$, where $f^\star_k$ is given by
Eq.~(\ref{ue7}), the potential (\ref{t1}) has two minima (double-well
potential). For a large force $|f|>f^\star_k$ the potential (\ref{t1})
has only one minimum (one-well potential) while for $f=f^\star_k$ the
potential is a one-well potential with a bending point.

The calculation of the DOS for the lower minimum in the two-well case can
be considered together with the one-well case ($k<0, f>f^\star$).
The position $x_0$ of the minimum can be found from the equation
\begin{equation}
  \label{hb9}
  Ax_0^3-|k|x_0=f .
\end{equation}
For $f>0$ and $f<0$  one should take, respectively, the positive and
negative root of this
cubic equation. In this minimum the oscillator has
a harmonic frequency
\begin{equation}
  \label{v5r}
  M\omega_{\rm new}^2 = -|k|+3Ax_0^2 .
\end{equation}
The density of states can be calculated from the Eq.~(\ref{des2}).
For $\omega\ll\omega_c$
\begin{equation}
  \label{mqc2}
g_{\rm II}(\omega) = 2n_0\Phi(0)\frac{M^2\omega^3}{\sqrt{3A}}
\int\limits_0^{M\omega^2/2}d|k|\frac{P\left[f(|k|)\right]}
{\sqrt{M\omega^2+|k|}}
\end{equation}
where according to Eqs.~(\ref{hb9}) and (\ref{v5r})
\begin{equation}
  \label{dse7}
  f(k)=Ax_0^3-|k|x_0=\frac{1}{3}
\sqrt{\frac{M\omega^2+|k|}{3A}}(M\omega^2-2|k|) .
\end{equation}

Comparing $f(k)$ with $f^\star_k$ given by Eq.~(\ref{ue7}) one can
see that the region of integration in Eq.~(\ref{mqc2}),
$0<|k|<M\omega^2/3$ corresponds to the case of one minimum
(one-well potential, $f(k)>f^\star_k$) and region
$M\omega^2/3<|k|<M\omega^2/2$ corresponds to the case of two
minima (double-well potential, $f(k)<f^\star_k$). Taking into
account Eq.~(\ref{t3}) for $P(f)$ and introducing a new variable
$t=\sqrt{1+|k|/M\omega^2}$ we finally get from Eq.~(\ref{mqc2})
\begin{equation}
\label{bpq1}
g_{\rm II}(\omega) = \frac{12}{\pi}n_0
M\Phi(0) \frac{\omega^2}{\omega^\star}
\left(\frac{\omega}{\omega^\star}\right)^2\int\limits_1^{\sqrt{3/2}}
\frac{\displaystyle dt}{\displaystyle
1+\left(\omega/\omega^\star\right)^6t^2(3-2t^2)^2} .
\end{equation}

As follows from this equation at small frequencies,
$\omega\ll\omega^\star$, $g_{\rm II}(\omega)\propto\omega^4$ and for moderately
high frequencies satisfying the inequality $\omega^\star\ll\omega\ll\omega_c$,
the integral in Eq.~(\ref{bpq1}) is equal to
$(\pi/12)(\omega^\star/\omega)^3$. Therefore in this case
$g_{\rm II}(\omega)=n_0M\Phi(0)\omega$ what coincides with
Eq.~(\ref{rf3}).

Combining results (\ref{bp1}) and (\ref{bpq1}) we get for the
total DOS at $T=0$
\begin{equation}
\label{gd3}
g_{\rm tot}(\omega)=g_{\rm I}(\omega)+g_{\rm II}(\omega)
= \frac{12}{\pi}n_0
M\Phi(0) \frac{\omega^2}{\omega^\star}
\left(\frac{\omega}{\omega^\star}\right)^2\int\limits_0^{\sqrt{3/2}}
\frac{\displaystyle dt}{\displaystyle
1+\left(\omega/\omega^\star\right)^6t^2(3-2t^2)^2} .
\end{equation}
We want to mention that Eq.(\ref{gd3}) differs from the
corresponding equation (22) of Ref.~\cite{GPS} not only by the
prefactor and the upper limit but also by the power of $(3-2t^2)$ in the
integrand. Thus we are correcting our error in Ref.~\cite{GPS}. Fortunately
this does only marginally alter the plots of the quoted paper, where the
analytical theory is compared to the results of simulation and
experiment, since the plots of the present function 
$g_{\rm tot}(\omega)$ and the one given in Ref.~\cite{GPS} differ
only slightly. 

For $\omega\ll\omega^\star$ the integral in Eq.~(\ref{gd3}) is equal
to $\sqrt{3/2}$ and we have 
\begin{equation}
  \label{ff3}
g_{\rm tot}(\omega) = \frac{12}{\pi}\sqrt{\frac{3}{2}}n_0M\Phi(0)
\omega\left(\frac{\omega}{\omega^\star}\right)^3 \propto \omega^4,
\quad \omega\ll\omega^\star .
\end{equation}
Taking into account Eq.~(\ref{wdef}) one can present the
density of states as function of energy $E=\hbar\omega$
(for $E\ll E^\star\equiv\hbar\omega^\star$) in the following way
\begin{equation}
  \label{ja2}
n(E) =  g_{\rm tot}(\omega)/\hbar = \frac{1}{\sqrt{2}}n_0\Phi(0)P(0)
\frac{M^{3/2}W^{5/2}}{\hbar^3}\left(\frac{E}{W}\right)^4 .
\end{equation}
This result for $n(E)$ can be compared with Eq.~(\ref{t8})
giving the density of states for TLS's. It is clear from this
comparison that for $E\gg W$ the density of states of HO's is much
bigger than the density of states for TLS's. For large frequencies
$\omega^\star\ll\omega\ll\omega_c$ the integral in Eq.~(\ref{gd3})
is equal to $(\pi/4)\cdot(\omega^\star/\omega)^3$ and we have
\begin{equation}
  \label{es2}
g_{\rm tot}(\omega) = 3n_0M\Phi(0)\omega , \qquad
\omega^\star\ll\omega\ll\omega_c   ,
\end{equation}
which coincides with Eq.~(\ref{fv1}).

\begin{figure}[htb]
\includegraphics[bb=330 553 595 770,angle=-180,totalheight=6cm,
keepaspectratio]{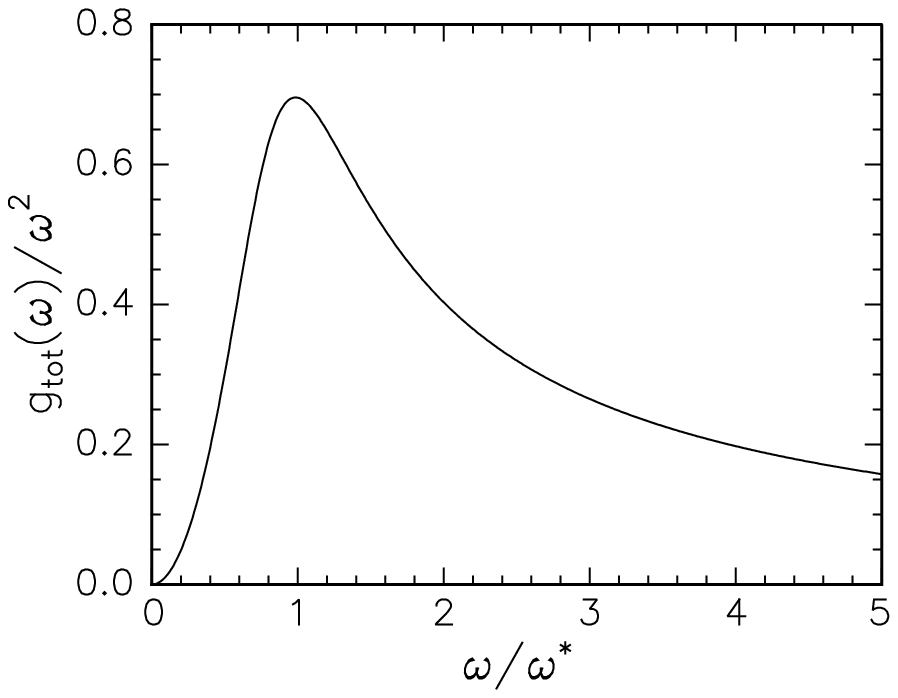}
\caption{The Boson peak: the reduced density of states $g_{\rm tot}(\omega)/\omega^2$
given by Eq.~(\ref{gd3}).}
\label{fig:bpeak}
\end{figure}

Since at low frequencies $\omega\ll\omega^\star$, the total DOS
$g_{\rm tot}(\omega)\propto\omega^4$ and at high frequencies
$\omega^*\ll\omega\ll\omega_c$, $g_{\rm tot}(\omega)\propto\omega$ we
have a peak in the reduced density of states
$g_{\rm tot}(\omega)/\omega^2$ at $\omega\approx\omega^\star$, the
Boson peak. In the Fig.~\ref{fig:bpeak} we plot the function
$g_{\rm tot}(\omega)/\omega^2$. We see from this figure that
$\omega_b\approx\omega^\star$.
%%%%%%%%% Picture 3   %%%%%%%%%%%%%%%%
%%%%%%%%% End Picture 3 %%%%%%%%%%%
This figure is valid only for the case $\omega^*\ll\omega_c$. 
If $\omega^*\simeq\omega_c$ ($g_0(\omega)\approx {\rm const}$) then 
$\omega_b\simeq\omega_c$ and at 
$\omega\gg\omega_c$, according to our previous results (see Ref.~\cite{GPS}, Eq.~(27)),
$g_{\rm tot}(\omega)\approx g_0(\omega)\approx {\rm const}$. In this case the right
wing of the Boson peak is determined by the initial density of states $g_0(\omega)$ and 
$g_{\rm tot}(\omega)/\omega^2 \propto 1/\omega^2$ 
(instead of $g_{\rm tot}/\omega^2\propto 1/\omega$ in the previous case).

The DOS for the higher minimum in the double-well potential
(\ref{t1}) is different. Though the thermal occupation number of this
minimum is smaller than the one of the lower minimum it can
contribute to the total DOS at finite temperatures. Starting from
Eq.(\ref{des2}) the position of the higher minimum can be obtained as
the smallest negative root or largest positive root of
Eq.~(\ref{hb9}) for $f>0$ and $f<0$, respectively.
The resulting DOS for $\omega\ll\omega_c$ is
\begin{equation}
  \label{xu8}
g_{\rm III}(\omega) = 2n_0\Phi(0)\frac{M^2\omega^3}{\sqrt{3A}}
\int\limits_{M\omega^2/2}^\infty d|k|\frac{P\left[f(|k|)\right]}
{\sqrt{M\omega^2+|k|}}
\end{equation}
with
\begin{equation}
  \label{xs7}
  f(k)=Ax_0^3-|k|x_0=\frac{1}{3}
\sqrt{\frac{M\omega^2+|k|}{3A}}(2|k|-M\omega^2) .
\end{equation}
Inserting the Lorentzian distribution for $P(f)$, Eq.~(\ref{t3}),
and introducing a new
variable $t=\sqrt{1+|k|/M\omega^2}$ we get for the DOS
\begin{equation}
\label{bb3}
g_{\rm III}(\omega) = \frac{12}{\pi}n_0
M\Phi(0) \frac{\omega^2}{\omega^\star}
\left(\frac{\omega}{\omega^\star}\right)^2\int
\limits_{\sqrt{3/2}}^\infty
\frac{\displaystyle dt}{\displaystyle
1+\left(\omega/\omega^\star\right)^6t^2(3-2t^2)^2} .
\end{equation}

At low frequencies, $\omega\ll\omega^\star$, the integral in
Eq.~(\ref{bb3}) is equal to $(\pi/3\cdot 2^{1/3})(\omega^\star/
\omega)$ and
\begin{equation}
  \label{hj3}
  g_{\rm III}(\omega) = 2^{5/3}n_0M\Phi(0)\omega^\star
\left(\frac{\omega}{\omega^\star} \right)^3 \propto\omega^3 ,
\quad \omega\ll\omega^\star .
\end{equation}
We see that at low frequencies the dependence of $g_{III}(\omega)$ 
(Eq.~(\ref{bb3})) differs from $\omega^4$ dependence, the DOS in
the higher minimum is proportional to $\omega^3$ in accordance
with Ref.~\cite{GC}. From Eqs.~(\ref{ff3}) and (\ref{hj3}) it
follows that $g_{\rm III}(\omega)/g_{\rm  II}(\omega)\approx
\omega^\star/\omega \gg 1$ and therefore, for
$\omega\ll\omega^\star$ and equal population of both minima the
DOS in the higher minimum is larger than the one in the lower
minimum. However, as we will see below, including the thermal
population factors for the two minima reverses this.

For high frequencies, $\omega\gg\omega^\star$, the integral in
Eq.~(\ref{bb3}) is  $(\pi/12)(\omega^\star/\omega)^3$ and
\begin{equation}
  \label{u6g}
  g_{\rm III}(\omega)=n_0M\Phi(0)\omega ,\quad \omega\gg\omega^\star
\end{equation}
which coincides with Eq.~(\ref{rf3}).

So far we disregarded the thermal population factor. Taking it into
account we get a temperature weighted DOS
\begin{equation}
  \label{xu9}
\tilde{g}_{\rm III}(\omega,T) = 2n_0\Phi(0)\frac{M^2\omega^3}{\sqrt{3A}}
\int\limits_{M\omega^2/2}^\infty d|k|\frac{P\left[f(|k|)\right]}
{\sqrt{M\omega^2+|k|}} \frac{1}{1+\exp(\Delta/T)}
\end{equation}
where $\Delta$ is the energy difference between the minima.

In the low-frequency case, $\omega\ll\omega^\star$, the integral in
Eq..~(\ref{xu9}) is a constant plus some {$\omega$}-dependent
correction. To estimate the constant, we set $\omega=0$ in the integral.
The higher minimum then turns into a bending point and
\begin{equation}
  \label{mj5}
  \tilde{g}_{\rm III}(\omega,T) = 2n_0\Phi(0)\frac{M^2\omega^3}
{\sqrt{3A}}\int\limits_0^\infty\frac{d|k|}{\sqrt{|k|}}
\frac{P(f^\star_k)}{1+\exp(\Delta/T)},
\end{equation}
where $\Delta = 3k^2/4A$ is the energy distance between positions of the
minimum and the bending point in the potential $U_{\rm eff}(x)$ for
$f=f^\star_k$ [see Eq.~(\ref{ue7})].

The result of the integration in Eq.~(\ref{mj5}) depends on which of
the two functions $P(f^\star_k)$ or $\left[1+\exp(\Delta/T)\right]^{-1}$
decays faster with $|k|$. The function $P(f^\star_k)$ decays with
a characteristic scale
\begin{equation}
  \label{nb4}
 |k|=k_f=3A^{1/3}(\delta f)^{2/3}/2^{2/3}
\end{equation}
(for $f^\star_k=\delta f$). The function
$\left[1+\exp(\Delta/T)\right]^{-1}$
decays with a characteristic scale $|k|=k_T=2\sqrt{AT/3}$. Both scales
become equal at the temperature $T=T^\star$, where $T^\star$ is given by
\begin{equation}
  \label{oz1}
  T^\star=\frac{27}{8\cdot 2^{1/3}}\frac{(\delta f)^{4/3}}{A^{1/3}}
=\frac{3}{8\cdot 2^{1/3}}\frac{M^2}{A}(\omega^\star)^4 =
\frac{3}{128\cdot 2^{1/3}}\hbar\omega^\star
\left(\frac{\hbar\omega^\star}{W}\right)^3.
\end{equation}
Estimates show that $T^\star$ is a rather large. For example
for $\hbar\omega^\star=40\,$K  and $W=4\,$K we have
$T^\star\approx 740\,$K.

Therefore, the low-temperatures case is more realistic. For $T\ll
T^\star$, one has $k_T\ll k_f$ so that $P(f^\star_k)=P(0)$ and
therefore from Eq.~(\ref{mj5})
\begin{equation}
  \label{pl7}
 \tilde{g}_{\rm III}(\omega,T) = \frac{2\sqrt{2}}{3^{3/4}}n_0\Phi(0)
P(0)\frac{M^2\omega^3}{A^{1/4}}T^{1/4}\int\limits_0^\infty
\frac{dy}{\sqrt{y}}\frac{1}{1+\displaystyle e^y} \propto
\omega^3 T^{1/4} .
\end{equation}
The last integral in this equation is equal $1.07\approx 1$. Therefore the 
equation can be rewritten in the form
\begin{equation}
  \label{sa1}
  \tilde{g}_{\rm III}(\omega,T) \approx \frac{4\sqrt{2}\cdot3^{3/4}}{\pi}
n_0\Phi(0)M\left(\frac{\omega}{\omega^\star} \right)^3
\frac{W}{\hbar}\left(\frac{T}{W} \right)^{1/4} .
\end{equation}

Now let us compare $g_{\rm tot}(\omega)$ and $\tilde{g}_{\rm
III}(\omega,T)$ for $\omega\ll\omega^\star$. Taking into account
Eqs.~(\ref{ff3}) and (\ref{sa1}) we have
\begin{equation}
  \label{ua1}
  \frac{g_{\rm tot}(\omega)}{\tilde{g}_{\rm III}(\omega,T)} \approx
\frac{\hbar\omega}{W}\left(\frac{W}{T} \right)^{1/4} .
\end{equation}
Thus for $T\ll W\left(\hbar\omega/W\right)^4$ we get
$g_{\rm tot}(\omega)\gg g_{\rm III}(\omega)$. In the opposite case
the contribution of the higher minimum to the DOS dominates.

\section{resonant scattering of phonons by HO'\lowercase{s}}

Taking Eq.~(\ref{int1}) for the coupling of the quasilocal oscillators to
the phonons we get
\begin{equation}
  \label{po6}
  l_{\rm res,HO}^{-1} = \frac{\pi\Lambda^2}{2M\rho v^3}\,g(\omega)
\end{equation}
where $l_{\rm res,HO}$ is the mean-free path of phonons due to
resonant scattering on quasilocal HO's with a density of states $g(\omega)$.
For low frequencies, below the Boson peak frequency, 
$\omega\ll\omega^\star$, we have from
Eq.~(\ref{ja2}) (see also Ref.~\cite{RB2})
\begin{equation}
  \label{fk5}
 l_{\rm res,HO}^{-1} = \frac{\pi}{8}\frac{C\omega}{v}
\left(\frac{\hbar\omega}{W}\right)^3\propto \omega^4
\end{equation}
where $C\simeq 10^{-3}\div 10^{-4}$ is the TLS's dimensionless parameter 
given by Eqs.~(\ref{yu6g}) and (\ref{d5}) (see also the estimate, 
Eq.~(\ref{d8})).
Its value is well known from the low temperature properties of glasses.
For high frequencies, above the Boson peak, in the interval 
$\omega^\star\ll\omega\ll\omega_c$, we have from
Eq.~(\ref{es2}), $g(\omega)=3n_0M\Phi(0)\omega$. As a result
\begin{equation}
  \label{ap2}
l_{\rm res,HO}^{-1} = \frac{3}{2}\pi\frac{\Lambda^2n_0}{\rho v^2}\,
\Phi(0)\,\frac{\omega}{v}\propto\omega.
\end{equation}

Let us compare the last quantity (proportional to $\omega$) with the 
inverse wave length of the phonons $\lambda^{-1}=\omega/2\pi v$. We have
the ratio
\begin{equation}
  \label{xt8}
  \frac{\lambda}{l_{\rm res,HO}} =
3\pi^2\,\frac{\Lambda^2n_0}{\rho v^2}\,\Phi(0).
\end{equation}
Using the estimates
\begin{equation}
  \label{7gb}
\Lambda^2n_0/\rho v^2\approx I\approx M\omega_c\omega_0,
\qquad \Phi(0)\approx F(k_c) = g_0(\omega_c)/2M\omega_c, 
\end{equation}
and $\omega_0\simeq 1/g_0(\omega_0)$ we have
\begin{equation}
  \label{pm5}
\frac{\lambda}{l_{\rm res,HO}}\approx\frac{3}{2}\pi^2\,
\frac{g_0(\omega_c)}{g_0(\omega_0)} \approx \frac{3}{2}\pi^2\, 
\left(\frac{\omega_b}{\omega_c}\right)^3,
\end{equation}
the last estimate follows from Eq.~(\ref{k}). 
Thus this ratio is a constant in the interval $\omega^* < \omega < \omega_c$ and  
depends only on the characteristic frequency $\omega_c\propto I$ and 
the behavior of the initial DOS $g_0(\omega)$. From the last equation it follows
that ratio $\lambda/l_{\rm res, HO}$ depends on the cube of the ratio of 
two important frequencies, the 
Boson peak frequency $\omega_b\approx\omega^*$ and the characteristic 
frequency $\omega_c$. Both of them can be measured on experiment 
(see Eq.~(\ref{st6r}) and Ref.~\cite{ParLaer}).

For the weak interaction $I$ which we consider in the paper 
$\omega_c\ll\omega_0$ and if for $\omega\to 0$ initial DOS $g_0(\omega)$ 
also goes to zero sufficiently rapidly, then $g_0(\omega_c)\ll g_0(\omega_0)$ and
$\lambda\ll l_{\rm res, HO}$, i.e. resonant phonon scattering is also weak. However,
due to big numerical coefficient in Eq.~(\ref{pm5}) in some realistic cases we
can have a strong phonon scattering.  
For example if the initial DOS $g_0(\omega)\propto\omega^2$ and $\omega_c\approx\omega_0/3$, then
\begin{equation}
  \label{qp7}
\lambda/l_{\rm res,HO} \approx \pi^2/6 \approx 1.64 .
\end{equation}
In this case criterion of Ioffe-Regel for the phonons ($l_{\rm res, HO} < \lambda$) is
approximately satisfied and we have strong phonon scattering above 
the Boson peak frequency in the interval $\omega^*\ll\omega\ll\omega_c$ 
(which is not too big since in this case $\omega_c\approx 2\omega^*$).

Another interesting case is a flat initial DOS, 
$g_0(\omega)\approx {\rm const}$, then 
$g_0(\omega_c)\simeq g_0(\omega_0)$ and 
as follows from Eq.~(\ref{k}) $\omega^*\simeq \omega_c$ and interval 
$[\omega^*,\omega_c]$ shrinks to one point (Boson peak frequency, 
$\omega_b\simeq\omega_c$) and at the Boson peak frequency we have
\begin{equation}
\frac{\lambda}{l_{\rm res,HO}}=\frac{3}{2}\pi^2 \approx 15  .
\label{ja5r}
\end{equation}
In this case the criterion of Ioffe-Regel is again satisfied 
and we have at the Boson peak frequency $\omega^*\simeq \omega_c$ 
the regime of {\em very strong resonant scattering} of phonons on quasilocal 
harmonic oscillators, independent of the strength of interaction $I$ 
(and $\omega_c$). However, it is necessary to stress that in this case 
the strong scattering takes place only in the vicinity of the Boson peak frequency 
$\omega_c$. 

At higher frequencies the initial DOS $g_0(\omega)\simeq {\rm const}$ and according to 
Eq.~(\ref{po6}) the resonance phonon mean free path is also constant and 
independent of frequency. However the phonon wave length $\lambda\propto 1/\omega$
decreases with frequency. Therefore the regime of the weak phonon scattering 
will recover again at higher frequencies $\omega \gg \omega_c$. 
Similar behavior was observed in Ref.~\cite{Sethna90} for resonant 
scattering of phonons on librational (quasilocal) modes in crystals.
We can extend the regime of strong scattering to well above the Boson peak frequency 
(up to Debye frequency $\omega_0$) only when the
initial DOS is a linear function of the frequency, $g_0(\omega)\propto\omega$ 
and interaction $I$ is not too small, $\omega_c > \omega_0/15$ (see Eq.~(\ref{pm5})). 

As we already mentioned, we will have a weak resonant scattering 
of phonons on quasilocal oscillators 
only when the interaction $I$ is sufficiently weak and the initial 
density of states $g_0(\omega)$ decreases to zero sufficiently fast 
with $\omega$, so that $g_0(\omega_c)\ll (2/3\pi^2)\, g_0(\omega_0)$. 
In such a case the mean free path of the phonons $l_{\rm res, HO}$ 
will be much larger than their wave length $\lambda$ in the whole 
frequency range. In this case phonons are well defined quasiparticles 
everywhere.

We give here also the relaxation time $\tau$ of a HO with
frequency $\omega$ due to the interaction with phonons. From
Eq.~(\ref{int1}) we get
\begin{equation}
  \label{gb5}
  \frac{1}{\tau} = \frac{\Lambda^2}{\rho v^2}
\frac{\omega^2}{4\pi Mv^3} = \frac{J^2\omega^2}{4\pi Mv^3} .
\end{equation}
Estimating
\begin{equation}
  \label{est1}
  \Lambda\simeq {\cal E}_0/a ,\quad \rho v^2\simeq {\cal E}_0/a^3 ,
\quad Mv^2\simeq {\cal E}_0 , \quad \hbar v/a \simeq \hbar\omega_0 ,
\end{equation}
where ${\cal E}_0\simeq 10\,$ eV is of the order of atomic energy,
$a\simeq1\,\mbox{\AA}$ is of the order of interatomic distance, and
$\omega_0$ is of the order of Debye frequency we get
\begin{equation}
  \label{est2}
  1/\omega\tau \simeq \omega/4\pi \omega_0
\ll 1 .
\end{equation}
Therefore HO's with $\omega\ll\omega_0$ are well defined objects.

\section{discussion}

In our previous papers~\cite{GPS,GPS1} we proposed a mechanism of the
Boson peak formation. The essence of the mechanism can be formulated
as follows. A vibrational instability of the weakly interacting QLV's
(stabilized by the anharmonicity) is responsible for the Boson peak
in glasses and other disordered systems. The instability occurs
below some frequency $\omega_c$ proportional to the strength of the interaction 
$I$ between low and high frequency oscillators. Whereas anharmonicity is
essential in creating the atomic structures supporting the Boson
peak, the vibrations forming the peak in the inelastic scattering
intensity or the reduced density of states are essentially harmonic.

The present paper extends these ideas. We show that such seemingly
unrelated phenomena in glasses (typical for the glassy state and usually
treated by separate unrelated models) as the formation of the two-level
systems and the Boson peak in the reduced density of low-frequency
vibrational states $g(\omega)/\omega^2$ can be explained by the same
physical mechanism, namely the {\em vibrational instability} of
weakly interacting soft harmonic vibrations. These can be seen as
localized vibrations with a bilinear interaction with the extended
modes, the sound waves. The resulting exact harmonic eigenmodes are
quasilocalized vibrations that have been observed in numerous
computer simulations and have been discussed extensively --- see
Ref.~\cite{GPS} and the references therein.

The instability, which as in all solids is controlled by the
anharmonicity, creates a new stable {\em universal} spectrum of harmonic
vibrations with the Boson peak feature as well as double-well
potentials with a wide distribution of the barriers heights that is
determined by the strength of the interaction $I$ between the oscillators. 
Depending on the barrier height (and temperature) these will lead to 
tunneling and relaxational transitions. To check for the consistency of our
theory we calculated the dimensionless parameter  
$C=\overline{P}\gamma^2/\rho v^2\approx 10^{-4}$ for the two-level systems in 
glasses which is observed in experiment~\cite{Pohl}. 
The smallness of this parameter is a longstanding puzzle. In our theory it
follows naturally. The physical reason for small value
of the parameter $C$ is that only a small fraction of all created TLS's can 
actually tunnel in realistic timescales. 

We show that the larger is the interaction $I$ between the original 
harmonic vibrations the smaller is parameter $C$. It reminds partly the 
ideas of Ref.~\cite{Coop} though we do not have here the 
frustrated strong interactions. We prove that for our simple model
weakly interacting oscillators $C=(W/\hbar\omega_c)^3\propto I^{-3}$. 
Here $W$ is an important characteristic energy in glasses 
of the order of a few Kelvin. The value of $C$ is independent of the
assumed initial DOS of HO's $g_0(\omega)$ and in this sense it is
universal. Varying the characteristic energy $W$ and interaction $I$ 
(i.e. the characteristic energy $\hbar\omega_c$) for different
glasses $C$ lies in the interval from $10^{-3}$ to $10^{-4}$.
However, we want to  stress that we are not free in the choice of these two parameters.
The energy $W$ is well known from experiments on specific heat~\cite{BGGS,IKP}, thermal 
conductivity~\cite{RB} and heat release~\cite{SP} in glasses. As for the 
characteristic frequency $\omega_c$ it should be of the order or larger 
than the Boson peak frequency $\omega_b$.

In the unified approach developed in the present
paper the densities of tunneling states and of excess vibrational
states at the Boson peak frequency are interrelated. Since the experimental
values of $C$ and $W$ are well known for many glasses we can use this
formula to get the important energy $\hbar\omega_c=WC^{-1/3}$ giving
us the onset of the vibrational instability region. For vitreous silica
$\hbar\omega_c\approx 60\,$ K falling perfectly into the Boson peak range. The same
holds for many other glasses~\cite{ParLaer}. It indicates
that the Boson peak is indeed placed inside the vibrational
instability range. 

It is instructive to compare the results of the present paper and of
Refs.~\cite{GPS,GPS1} with our earlier paper~\cite{GPPS} where we
also discuss the possible origin of the Boson peak. In this paper we
consider low-energy Raman scattering in glasses. As in the present
paper and in Refs.~\cite{GPS,GPS1} we assume that the scattering and
the energy transfer are due to the interaction of the light with the
soft potentials in glasses. The density of states of the quasilocalized
HO's, according to Ref.~\cite{GPPS}, is proportional to $\omega^4$ for
low frequencies and to $\omega$ for high frequencies. This behavior
qualitatively resembles the one obtained in Ref.~\cite{GPS}.
However, the phenomenon of vibrational instability was disregarded
in~\cite{GPPS} and the discussion of the Boson peak was necessarily
somewhat qualitative. Considering the vibrational instability, puts
the theory on a more quantitative level and, for instance, allows the
determination of the shape of the Boson peak.

Let us now compare the results of our paper with previously published 
important class of models of the Boson peak~\cite{SDG,TLNE,KRB,GMPV}. 
Experiment has shown that the Boson peak is formed by largely harmonic 
vibrations. Therefore, in all these models the authors have considered 
an Hamiltonian of the form
\begin{equation}
U_{\rm tot}(x_1,x_2,...x_n) = \frac{1}{2}\sum_{i,j\ne i}k_{ij}(x_i-x_j)^2 
\label{4rs3}
\end{equation}
with randomly distributed  quasielastic constants $k_{ij}$. 
Since this potential energy is purely harmonic we call all such 
models of the Boson peak  {\em harmonic random matrix} (HRM) models.
The main difference between the quoted four HRM models is in the different 
distributions of the quasielastic constants $k_{ij}$.

If all quasielastic constants are positive, $k_{ij}>0$, 
then the corresponding dynamical matrix (Hessian) 
is positive-definite and therefore all the eigenvalues are obviously 
positive as well, $\omega_i^2>0$ ($i=1,2,...n$) excluding those zeroes which come 
from the translational and rotational invariances. In such a case the system is 
{\em mechanically stable}. As was shown in~\cite{SDG,TLNE,GMPV} 
the system remains to be stable even when some (rather small) fraction of the 
quasielastic constants $k_{ij}$ in Eq.~(\ref{4rs3}) is negative 
(and small enough). Increasing the fraction of negative $k_{ij}$ (or their 
absolute values) authors have approached the mechanical stability threshold 
(when the first negative $\omega^2$ has appears in the spectrum). 

The reduced density of states $g(\omega)/\omega^2$ for Hamiltonian (\ref{4rs3})
usually has a maximum at some frequency $\omega_{\rm max}$ for typical values
of $k_{ij}>0$. Changing the parameters of the distribution function $P(k_{ij})$ 
one can shift this maximum to higher or to low frequencies. 
The last case was the main goal of the papers~\cite{SDG,TLNE,KRB,GMPV}.
The biggest red shift of the maximum 
has been achieved approaching the mechanical stability threshold.
In the first two papers~\cite{SDG,TLNE} the original maximum was due to 
Van-Hove singularity of the crystalline DOS. Atoms in these papers were placed 
on a perfect cubic lattice. As a result the Boson peak has been ascribed
to the lowest Van Hove singularity shifted due to disorder. 
In another two papers~\cite{KRB,GMPV} atoms were distributed randomly in 3-d space 
(so-called Euclidean Random Matrix models) and quasielastic 
constants $k_{ij}$ depend exponentially on the interatomic distances 
$|{\bf r}_i-{\bf r}_j|$. Therefore the distribution function of quasielastic constants
in these two cases $P(k)\propto 1/k$ and has a singularity for $k\to 0$.  
Thereby the portion of small $k$ was increased 
compared to Gaussian and box distributions used in the papers~\cite{SDG,TLNE}. 
Obviously the red shift of the original peak in 
$g(\omega)/\omega^2$ was much more pronounced in these latter two 
cases~\cite{KRB,GMPV}. 

Our approach, which continues from our previous papers~\cite{GPS,GPS1}
differs essentially in two ways. First, we postulate that 
the excess in vibrational
modes originates from quasi-localized vibrations. Their existence 
has been shown in numerous simulations of different types of materials. 
Such modes can be described as local modes (cores) which weakly interact 
bilinearly with the extended modes (sound waves) and thus with each other. 
The exact harmonic eigenvectors are of course extended as in 
HRM models. Secondly, we do not invoke special distributions for the elements 
of the dynamical matrix to avoid unstable vibrations but, on the contrary, 
show that the generic instability, when controlled
by the anharmonicity which is present in all real systems, automatically gives both 
the TLS and the Boson peak (with a universal shape) without any further assumptions. 

The essence of the mechanism can be formulated as follows. The randomly distributed 
weakly interacting QLV become unstable at low frequencies in harmonic approximation. 
This is the equivalent to the instability in the general HRM models. Instead of 
assumptions on distribution functions of interactions $I_{ij}$ we use the always 
present anharmonicity as the stabilizing factor. The previous vibrational 
instability of the weakly interacting QLV's thus becomes responsible for
the Boson peak and TLS's in glasses and other disordered systems. Whereas anharmonicity
is essential in creating the atomic structures supporting the Boson peak,
the vibrations forming the peak in the inelastic scattering intensity or the reduced
density of states are essentially harmonic. Comparing with random matrix models our
Eq.~(\ref{potgen}) without the anharmonicity term would correspond to 
the case of the unstable random matrix whereas the result of Eq.~(\ref{1}) correspond
to the stabilized case. And, importantly, the anharmonicity strength $A$
(thanks to mirror transformation) does not enter in the expression for the 
renormalized frequency. So, the anharmonicity reconstructs the spectrum but the final result is 
independent of the strength of anharmonicity. The advantage
of our approach is that the stabilization is not a result of an additional assumption
but is a benefit of the vibrational instability + anharmonicity which is haunting the alternative
approach.  

Summarizing briefly we can say that the Boson peak in 
papers~\cite{SDG,TLNE,KRB,GMPV} was obtained by a purely harmonic ansatz 
inside the mechanically {\em stable}
region of their Hamiltonians. The position and the form of the peak depend strongly 
on the distribution function of quasielastic constants 
$P(k_{ij})$. In our approach the Boson peak
is built inside the mechanically {\em unstable} region of the harmonic potential
parameters. Therefore, the role of anharmonicity as stabilizing factor 
is crucial. But the form of the Boson peak appears to be universal and independent
of the initial assumptions about interaction $I_{ij}$ or the initial distribution 
function $g_0(\omega)$ or of the anharmonicity strengths $A_i$. 
Relating TLS's and Boson peak parameters in our theory we are able to show 
that the instability crossover frequency $\omega_c$ given by 
Eq.~(\ref{s6e4}) lies in the Boson peak region.  
Due to weakness of the interaction $I$  the universal reconstruction of the 
spectrum in our theory takes place in the low frequency range only,  
leaving high frequency range nearly unchanged. This is different in the 
HRM models~\cite{SDG,TLNE,KRB,GMPV} where the whole spectral range is 
completely reconstructed in the course of the change of the distribution 
function $P(k_{ij})$. 

As mentioned before in our theory the estimate for the dimensionless 
tunneling strength $C$ emerges in a naturally  (see Eq.~(\ref{d8})). 
Varying the characteristic energy $W$ and the interaction strength $I$ 
for different glasses $C$ falls into the 
interval between 10$^{-3}$ and 10$^{-4}$.
We want to mention that other explanations of the
smallness of the constant $C$ have been proposed.  Burin and
Kagan~\cite{BK} predict a number of universal properties of amorphous
solids, including the small values of the tunneling strength $C$, due
to a special form of interaction of such defect centers with internal
degrees of freedom like the TLS's. It is important that the
interaction between such centers falls off with distance $r$ as
$1/r^3$. In this case effects of correlations between many TLS's ({\em
dipole gap} effects) might be important at sufficiently low temperatures.

In our paper we have neglected all such many particle correlation
effects between tunneling TLS's which could further reduce (or stabilize) the value
of $C$. One should, however, keep in mind that the $1/r^3$ law of
interaction is valid as far as the effects of retardation play no
role. In dielectric glasses these effects are determined by the sound
velocity. At sufficiently large distances they could play a role that
would result in the variation of the interaction law. We believe that
the role of the retardation effects deserves a special investigation.

Another important difference between our and Burin and Kagan approaches is 
that the parameter $C$ in our theory is a quantum 
mechanical quantity, $C\propto\hbar$ 
(see Eq.~(\ref{tr3})). It disappears in the classical limit $\hbar\to 0$.
In other words the smallness of $C$ is directly related to the smallness of the 
quantum mechanical
probability for particles tunneling through high barriers in a glass 
(see also Eq.~(\ref{pz3})). Differently, the dipole gap effect  is 
purely classical since it is based on the classical 
dipole-dipole interaction between
TLS's. Therefore, it would be very interesting to elucidate which of two 
mechanisms (or both) dominates and is responsible for the small value of 
parameter $C$ in glasses. In particular it would be very interesting to calculate
the relaxation times for many particle correlations effects to build the 
dipole gap. 

An analysis of the low temperature properties of glasses along the
lines of the soft potential model based on a numeric simulation of a
Lennard-Jones glass was presented by Heuer and Silbey~\cite{heuer:96}.
They numerically searched for the energy minima of the glass and 
constructed double-well potentials for close minima. By extrapolation
to small values they were able to extract the distribution functions
for the soft potential parameters. These potentials correspond to our
potentials after inclusion of the interaction between the HO's. The
present theory is in agreement with their simulation results.

The theory presented in this paper deals with the effects of soft modes 
produced by disorder which can be expected to have a broad frequency
distribution. In the literature the term Boson peak is rather loosely
defined. Often it is used for any low frequency maximum in the reduced
DOS. In particular in plastic crystals, see eg. Ref.~\cite{brand:02}, 
soft HO are present even before disorder. Consequently disorder only
broadens their sharp DOS. Depending on the strength of this broadening
our theory will apply more or less to such cases. The same applies to
TLS which also can be present before disorder.

In summary, we have shown that the same physical mechanism
is fundamental for such seemingly different phenomena as formation of
the two-level systems in glasses and the Boson peak in the reduced
density of low-frequency vibrational states $g(\omega)/\omega^2$. In
this way two of the most fundamental properties of glasses are
interconnected.

\acknowledgments
V. L. G and D. A. P. want to acknowledge the hospitality as well as
financial support by the Forschungszentrum J\"ulich where part of
this work was done. D.A.P. wants also to acknowledge the hospitality and 
financial support of the Max-Planck-Institut f\"ur
Physik komplexer Systeme, Dresden where part of this work was done.

\end{document}